\documentclass{IEEEtran}
\usepackage[latin9]{inputenc}
\usepackage{graphicx,color,url}
\usepackage{xr}
\usepackage[numbers]{natbib}
\usepackage{authblk}


\begin{document}

\title{Multiple Signal Classification Algorithm for super-resolution
fluorescence microscopy}

\author[1]{Krishna Agarwal$^{*,}$\thanks{$^{*,}$uthkrishth@gmail.com}}
\author[2]{Radek Mach\'{a}\v{n}}

\affil[1]{Singapore-MIT Alliance for Research and Technology and }
\affil[2]{Department of Biological Sciences and Centre for BioImaging Sciences, National University of Singapore}
\maketitle

\begin{abstract}
Super-resolution microscopy is providing unprecedented insights into biology by resolving details much below the diffraction limit. State-of-the-art Single Molecule Localization Microscopy (SMLM) techniques for super-resolution are restricted by long acquisition and computational times, or the need of special fluorophores or chemical environments. Here, we propose a novel statistical super-resolution technique of wide-field fluorescence microscopy called MUltiple SIgnal Classification ALgorithm (MUSICAL) which has several advantages over SMLM techniques. MUSICAL provides resolution down to at least 50 nm, has low requirements on number of frames and excitation power and works even at high fluorophore concentrations. Further, it works with any fluorophore that exhibits blinking on the time scale of the recording. We compare imaging results of MUSICAL with SMLM and four contemporary statistical super-resolution methods for experiments of in-vitro actin filaments and datasets provided by independent research groups. Results show comparable or superior performance of MUSICAL. We also demonstrate super-resolution at time scales of 245 ms (using 49 frames at acquisition rate of 200 frames per second) in samples of live-cell microtubules and live-cell actin filaments.
\end{abstract}

Super-resolution fluorescence microscopy techniques aim at resolving details smaller than the Abbe diffraction limit of $\frac{\lambda}{2{\rm NA}}$, where $\lambda$ is the wavelength of the fluorescence emission and ${\rm NA}$ is the numerical aperture of the microscope objective. Most of these techniques use the blinking phenomenon where fluorophores switch between a bright (fluorescent) state and a long-lived dark state. A series of images is recorded over time. Each image has different intensity distribution because a different set of fluorophores were in the bright state during each image acquisition. The temporal information contained in the series is, then, used to construct a final image with improved spatial resolution. Single molecule localization microscopy (SMLM) techniques such as stochastic optical reconstruction microscopy (STORM) or photo-activated localization microscopy (PALM) are popular super-resolution techniques owing to their simplicity, few (if any) special requirements on instrumentation, and impressive resolution of approximately 20 nm~\cite{betzig2006imaging,hess2006ultra,rust2006sub}. However, they require that the fluorophores exhibit long dark states so that only a small subset of optically separable fluorophores are in the bright state in each frame of the image stack. This translates into requirements of long acquisition times and of chemical environment promoting long dark states and impeding bleaching.

The limitations of SMLM have motivated development of  techniques that rely on statistical independence of blinking of individual fluorophores rather than on long dark states~\cite{small2014fluorophore}. Such techniques include super-resolution optical fluctuations imaging (SOFI~\cite{dertinger2009fast}), Bayesian analysis of blinking and bleaching (3B~\cite{cox2012bayesian}), and entropy-based super-resolution imaging (ESI~\cite{ESI}). Although they relax the requirements of SMLM, they do not reach resolution achievable by localization microscopy (approximately 110\,nm for SOFI~\cite{geissbuehler2014live}, 80\,nm for ESI~\cite{ESI}, and 50\,nm for 3B~\cite{cox2012bayesian}) and they possess limitations of their own. For example, SOFI uses cumulants of the fluorescence blinking to enhance resolution; since cumulants of orders higher than 6 are prone to shot noise and do not have good approximations, the practically achievable resolution improvement is limited to the factor of $\sqrt{6}$~\cite{dertinger2009fast}. 3B uses a Markov process for modeling the blinking and bleaching of the fluorophores and an expectation maximization approach to determine the likelihood of an emitter being present at a given location. This approach is computationally intensive and its convergence to global minimum is not guaranteed.

Here, we propose a novel algorithm utilizing fluorescence blinking to enhance spatial resolution. The algorithm, called MUltiple SIgnal Classification ALgorithm (MUSICAL), achieves super-resolution by exploiting the eigenimages of the image stack which statistically represent its prominent structures and, then, applying the knowledge of the point spread function (PSF) of the imaging system to localize the structures to super-resolution scales. Like SOFI or 3B and other related techniques, MUSICAL requires neither special instrumentation nor special fluorophores. The sole requirement is statistically independent blinking of individual emitters. We tested MUSICAL on images of actin filaments and compared it with STORM, showing that both techniques give comparable resolution enhancements. We also compared MUSICAL with 3B, SOFI, ESI, and deconSTORM \cite{mukamel2012statistical} on datasets independently measured and provided by other super-resolution research groups \cite{EPFL_SMLM,3B} and show comparable or superior performance of MUSICAL. We also demonstrate that MUSICAL performs well in situations where STORM fails due to high density of fluorophores. Further, we show that MUSICAL can be used for live-cell fast imaging ($\sim$49 frames amounting to a total acquisition time of less than 250 ms) of cells using standard green fluorescent protein (GFP).

The idea of MUSICAL is inspired from MUltiple SIgnal Classification
(MUSIC) used in acoustics \cite{gruber2004time}, radar signal processing
\cite{schmidt1986multiple}, and electromagnetic imaging \cite{chen2008music}
for finding the contrast sources created due to scattering and contributing
to the measured signal. However, MUSICAL differs from MUSIC because
the emitters in fluorescence microscopy behave differently from the
contrast sources encountered in scattering. Firstly, the fluorophores exhibit
intermittent emission when exposed to continuous excitation, the intermittence
patterns of any two fluorophores being uncorrelated. Secondly, the
information of a molecule is concentrated in a small region defined
by the PSF in the image plane. Thus, there is a region of confidence
for the likely position of each molecule; pixels beyond that region
contribute only additional noise and almost no usable information in the
context of localizing molecules in the given region. Thirdly, MUSIC in its traditional
form needs a higher number of receivers (pixels in the camera) and measurements
(frames) than the number of contrast sources (fluorescent emitters).
Thus, when imaging an area with a high density of fluorophores, the
required number of frames may easily become impractically large.

MUSICAL overcomes those issues of MUSIC by adopting a sliding soft
window and stitching approach. MUSIC is applied only on a small part
of the image (the soft window) at a time. The soft window is, then,
scanned over the whole image and the individual MUSIC reconstructions
are stitched together to form the MUSICAL image. See Fig. \ref{fig:MUSICAL} for the flowchart of MUSICAL. Since the area corresponding to the soft window contains only a fraction of the total number of fluorophores in the imaged area, the required number of frames in
the stack is considerably reduced. Besides, this approach also suppresses
the problem of noise contribution from pixels further away from a
given emitter location.

Using the selected soft window of the image stack, MUSIC first computes
the eigenimages of the soft window through singular value decomposition.
Each eigenimage is associated to a singular value and represents a
particular pattern found in the image stack. Statistically, large
singular value indicates that the pattern of the corresponding eigenimage
is a prominent pattern in the image stack. On the other hand, statistically
less likely patterns, which include noise patterns, are characterized by small singular values. Thus,
the eigenimages can be separated into two sets (or subspaces) using a threshold value $\sigma_0$, where eigenimages with singular values more than or equal to $\sigma_0$ form the signal subspace (or the range) and the remaining eigenimages form the null subspace. See Fig. \ref{fig:SynEx2NoiseLess} for an example showing the effect of the values of $\sigma_0$ on MUSICAL images.

Then, the projections of the PSF (Fig. \ref{fig:G_PSF}) at a given test
point on the range and the null subspace are determined, which we
denote as $d_{{\rm PR}}(\vec{r}_{{\rm test}})$ and $d_{{\rm PN}}(\vec{r}_{{\rm test}})$
respectively. The projection $d_{{\rm PR}}(\vec{r}_{{\rm test}})$
indicates whether the PSF at the test point is related to the patterns
represented by the range, similarly for $d_{{\rm PN}}(\vec{r}_{{\rm test}})$.
If the test point indeed belongs to either the fluorophore locations
or the structure represented by them, then $d_{{\rm PN}}(\vec{r}_{{\rm test}})$
is close to zero, it is non-zero otherwise. See Fig. \ref{fig:EigenImages} for an example of eigenimages and projections of PSF on the eigenimages.  This property of $d_{{\rm PN}}(\vec{r}_{{\rm test}})$
is used to compute an indicator function as follows
\[
f(\vec{r}'_{{\rm test}})=\Bigg({\frac{{d_{{\rm PR}}(\vec{r}'_{{\rm test}})}}{{d_{{\rm PN}}(\vec{r}'_{{\rm test}})}}}\Bigg)^{\alpha}.
\]
such that the value of the indicator function is very large
at the point a fluorophore is present and small at the point where no
fluorophore is present. In the original form of MUSIC, as used in acoustic
and electromagnetic imaging, only the distance from the null space
$d_{{\rm PN}}(\vec{r}_{{\rm test}})$ is used and $\alpha$ is often
set as 1. However, the indicator function of MUSIC is not suitable for
the sliding window approach, as shown in the supplement using a synthetic
example (Fig. \ref{fig:SynEx2}). The modified indicator function of MUSICAL is amenable to the
sliding window approach since the inclusion of $d_{{\rm PR}}(\vec{r}_{{\rm test}})$
in the numerator automatically weighs the result of each window so
that stitching of the results for sliding windows can be simplified
as sum of the indicator functions of all the sliding windows
covering a test point. At the same time, the use of $\alpha$ allows for better
resolution by non-linearly scaling the indicator function at closely
located test points (see Fig. \ref{fig:SynEx1} for an example and section \ref{sec:alpha} of the supplement for a discussion on the role of the parameter $\alpha$). More algorithmic and implementation details are provided in sections \ref{sec:MUSICAL} and \ref{sec:parameters}, respectively, of the supplement.

\section*{Results}

\begin{figure*}[ht!]
\centering \includegraphics[width=1\linewidth]{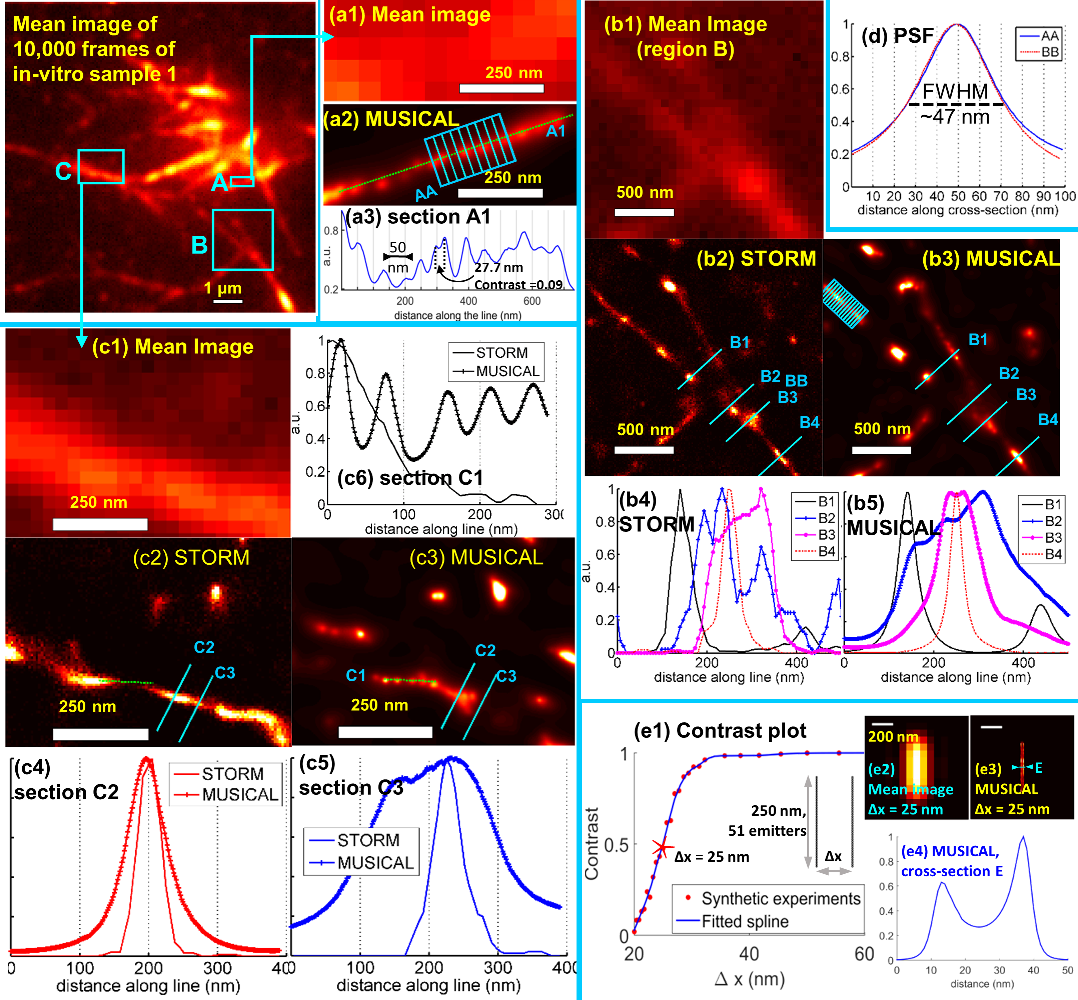}\\
 \caption{Results for different regions of in-vitro sample 1 are shown here. The mean image of 10,000 frames is shown in top-left corner. (a-c) correspond
to three regions A-C of the sample shown in the mean image. (a1-a2) show the mean image and
result of MUSICAL for region A. (a3) shows the profile of section A1
shown in (a2). (b1-b3) show the mean image, result of STORM, and result of MUSICAL for region B. (b4) shows the STORM profiles of sections B1-B4
shown in (b2). (b5) shows the MUSICAL profiles of sections B1-B4 shown
in (b3). (c1-c3) show the mean image, STORM result, and MUSICAL result
for region C. (c4) shows the profiles of section C2 from the STORM
and the MUSICAL result in (c2,c3) respectively. (c5) shows the profiles
of section C3 from the STORM and the MUSICAL results in (c2,c3) respectively.
(c6) shows the profile of section C1 shown in (c3). (d) shows the
average point spread functions derived from groups AA and BB of line
segments shown in (a2) and (b3), respectively. (e) shows the estimated contrast between two lines of emitter separated by distance $\Delta x$ computed using synthetic dataset SynPair. The contrast plot and the geometry of SynPair is shown in (e1). The red colored star shown on the plot in (e1) corresponds to $\Delta_x=25$ nm and has a contrast of 0.48. The mean image for the image stack and MUSICAL result for the SynPair data with $\Delta_x=25$ nm are shown in (e2,e3), respectively. The intensity at section E shown in (e3) is plotted in (e4).}
\label{fig:main_result}
\end{figure*}

\begin{figure*}[ht!]
\centering \includegraphics[width=1\linewidth]{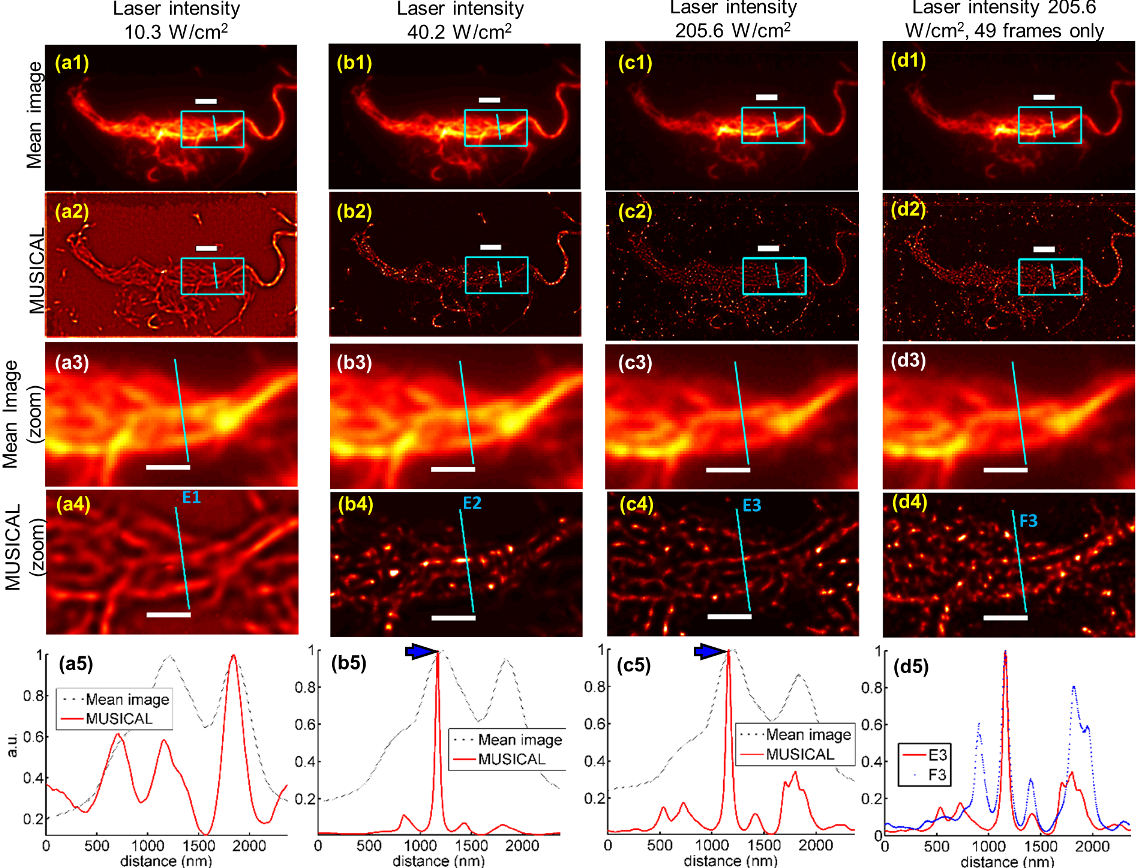}
\caption{MUSICAL results for in-vitro sample 2 are presented here. The first, second, and third columns correspond to the image stacks acquired using laser powers 10.3 W/cm$^2$, 40.2 W/cm$^2$, and 205.6 W/cm$^2$, respectively. The fourth column shows result of MUSICAL using 49 frames only from the image stack acquired with 205.6 W/cm$^2$ excitation power. (a5,b5,c5) compare the intensity profiles of mean image and MUSICAL at the sections E1, E2, and E3, respectively. (d5) compares MUSICAL result at E3 using entire stack of 10,000 images with MUSICAL result at F3 using 49 frames only. Scale bars: 1st and 2nd rows, 2 $\rm \mu$m; 3rd and 4th rows, 1 $\rm \mu$m.\vspace{-2mm}}
\label{fig:nonSTORM}
\end{figure*}

\begin{figure*}[ht!]
  \centering
  \includegraphics[width=\linewidth]{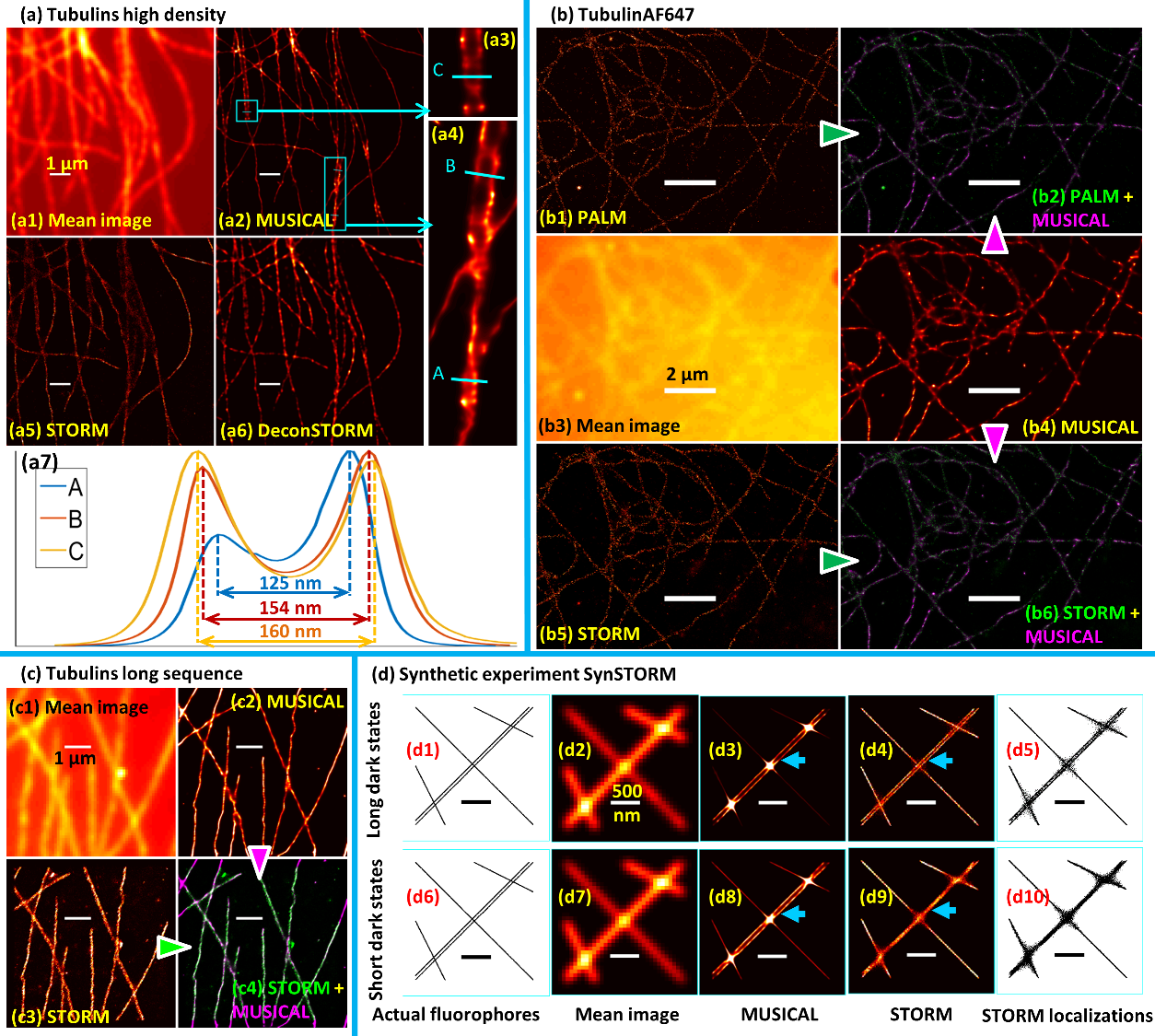}\\
  \caption{Results on Data-SMLM datasets and synthetic example SynSTORM for comparison with SMLM techniques are presented here. (a) Tubulins high density (500 frames), (b) Tubulins long sequence (15000 frames), (c) TubulinAF647 (9990 frames), (d) SynSTORM. (a1,a2,a5,a6) show the mean image, MUSICAL image, STORM image, and deconSTORM image, respectively. (a3,a5) show zoom-ins of two regions of (a2). (a7) plots the intensity at cross-sections A-C shown in (a3,a4) and shows that MUSICAL consistently resolves microtubules placed closer than the diffraction limit of 265 nm. (b1,b3,b4,b5) show the PALM result, mean image, MUSICAL image, and STORM result. PALM and STORM results are taken from \cite{EPFL_SMLM}. (b2,b6) show MUSICAL image (in green) overlaid with PALM image and STORM image respectively. (c1-c3) show the mean image, STORM image, and MUSICAL image, respectively. (c4) shows MUSICAL result overlaid with STORM result. (d1-d5) correspond to fluorophores with long dark states while (d6-d10) correspond to fluorophores with shorter dark states. (d1,d6) show actual emitter profiles. (d2,d7) show mean images of the image stacks. (d3,d8) show MUSICAL result. (d4,d9) show STORM images and (d5,d10) show the points localized by STORM.}\label{fig:EPFL}
\end{figure*}

\begin{figure*}[ht!]
\centering \includegraphics[width=1\linewidth]{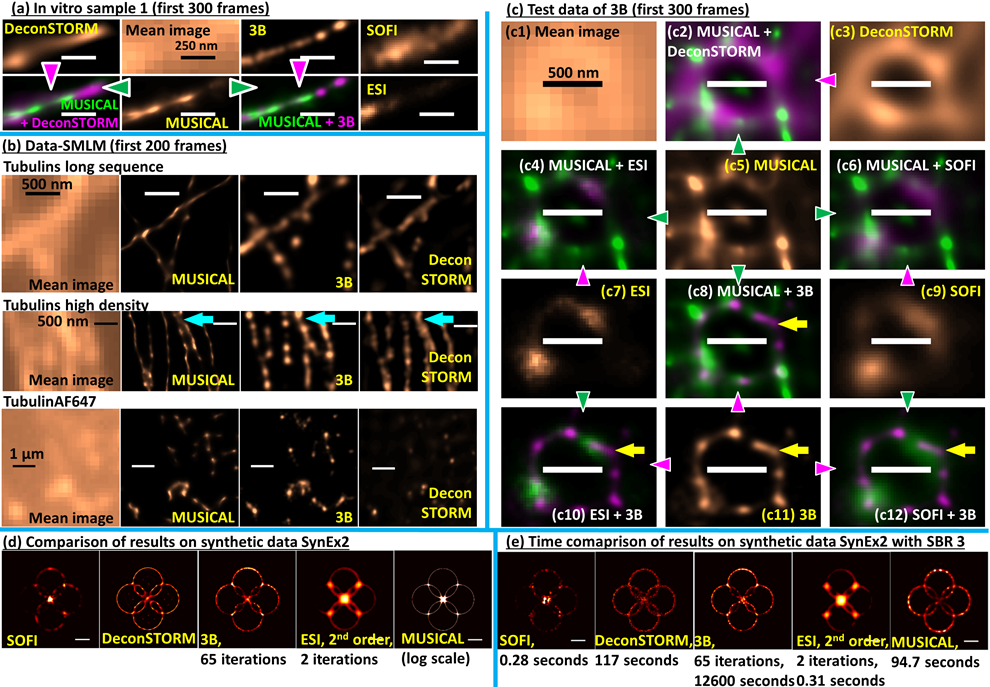}
\vspace{-7mm}
\caption{Comparison of MUSICAL with other super-resolution methods that perform statistical analysis of blinking statistics rather than SMLM are presented here. (a) shows comparison for region A of in-vitro sample 1. Overlay of MUSICAL image with 3B and deconSTORM is given for comparison in addition to their individual results. (b) shown comparison of MUSICAL results with 3B and deconSTORM for the three data in Data-SMLM. MUSICAL can resolve two closely lying microtubules (cyan colored arrow) which are unresolvable by 3B and deconSTORM. (c) presents comparison results for the test data of 3B. Overlay of different pairs of methods are given for the ease of comparison. (d) shows comparison results for synthetic example SynEx2 which is noise-free. (e) shows comparison results for synthetic example SynEx2 with Poisson noise statistics and signal to background ratio 3. Computation time for each method is provided for SynEx2SBR3 below its result.}
\label{fig:Comparison}
\end{figure*}

We present results of MUSICAL for in-vitro experiments, in-vitro test data available online \cite{EPFL_SMLM,3B}, and live-cell experiments. In-vitro experiments were performed on actin filaments tagged with Phalloidin-Atto 565 dye and serve to characterize the performance of MUSICAL. Out of three sample, results of samples 1 and 2 are included in Fig. \ref{fig:main_result} and Fig. \ref{fig:nonSTORM}. Experimental data provided as single molecule localization test dataset \cite{EPFL_SMLM} and the test data provided on the 3B project webpage \cite{3B} are used for comparison with other contemporary methods. Details about these datasets can be found in their respective websites. We refer to the former as Data-SMLM. It comprises of three data, namely tubulins long sequence (15,000 frames), tubulins high density (500 frames), tubulinAF647 (9990 frames). Tubulins high density data is further used to demonstrate resolution of two microtubules using MUSICAL. The results for Data-SMLM are included in Fig. \ref{fig:EPFL} and Fig. \ref{fig:Comparison}(c) and results for the test data of 3B are included in Fig. \ref{fig:Comparison}. Results of MUSICAL on live-cell microtubules sample 1 are included in Fig. \ref{fig:LiveMicrotubules}. Results on in-vitro actin sample 3 (Fig. \ref{fig:power}), time-lapse results for live-cell microtubules sample 1 (Fig. \ref{fig:LiveCellsSample1TimeLapse}), results for live cell microtubules sample 2 (Fig. \ref{fig:LiveCellsSample2}), and live-cell cortical cytoskeletal actin (Fig. \ref{fig:LiveCellActin})) are included in the supplement. The details of all the samples are provided in Table \ref{tab:samples}.

We breakdown the results into the following studies: characterization of the point spread function and resolution of MUSICAL, performance of MUSICAL for non-sparse blinking, minimum number of frames required for MUSICAL, excitation power and MUSICAL, comparison of MUSICAL with single molecule localization approaches, comparison of MUSICAL with other super-resolution techniques, and live cell results. These studies are presented below.

\vspace{2mm}
\noindent\textbf{Point spread function (PSF) and resolution of MUSICAL \textemdash{}} In-vitro sample 1 is used for characterization of the point spread function (PSF) of MUSICAL. In Fig. \ref{fig:main_result}(a2), group of line segments {AA} is used to obtain the experimental point spread function of MUSICAL. The lines are separated by 26 nm. The maxima of the profiles of all the lines are aligned and then their average profile is computed as representative of the PSF of MUSICAL, which is shown in Fig. \ref{fig:main_result}(d). The full width at half maximum (FWHM) of the average profile is 47.4 nm. We also determined FWHMs of all the individual profiles; they were 47.3 $\pm$ 7.4 nm (mean $\pm$ standard deviation). We repeated the same procedure for another group of lines BB shown in Fig. \ref{fig:main_result}(b3). The lines are separated by 65 nm here. The FWHM of the average profile for group BB is 46.6 nm. The FWHMs of the individual profiles were 48.5 $\pm$ 8.7 nm (mean $\pm$ standard deviation).

In order to study practically achievable resolution, we plot the contrast of MUSICAL images of two synthetic lines of emitters as a function of the distance $\Delta x$ between them in Fig. \ref{fig:main_result}(e1). The contrast is defined as $\frac{I_{\rm max}-I_{\rm min}}{I_{\rm max}+I_{\rm min}}$ where $I_{\rm max}$ is the maximum intensity among the two consecutive maxima and $I_{\rm min}$ is the intensity of the minimum between them. The details of the synthetic experiment, referred to as SynPairDelX, are discussed in section \ref{sec:resolution} of the supplement. As an example, we show in Fig. \ref{fig:main_result}(e3) the MUSICAL result for SynPair data corresponding to $\Delta x = 25$ nm, for which MUSICAL results into a contrast of 0.48. The intensity at section E shown using cyan line in Fig. \ref{fig:main_result}(e3) is shown in Fig. \ref{fig:main_result}(e4). It shows that the maxima are clearly separated with a distance of 25 nm between them. No other statistical method could resolve these two lines. STORM localized only three emitters for this data.

We observed periodicity along the actin filaments in the MUSICAL images and the minimum distance between the two peaks on actin filament to be 27.7 nm, as seen in Fig. \ref{fig:main_result}(a3). Further, we consider the tubulins high density data from Data-SMLM dataset which has high density of emitters such that blinking is non-sparse. Yet, MUSICAL clearly illustrates the ability of resolving the microtubules separated by 125 nm with a contrast value of 0.39, as seen for the section A in Fig. \ref{fig:EPFL}(a).

To study further the observed periodicity in the actin filaments, we plot the intensity of MUSICAL result across the length of sections of two actin filaments A1 and C1 in Fig. \ref{fig:main_result}(a3,c6), respectively. Fourier analysis of the profiles along A1 revealed a prominent peak at the spatial frequency corresponding to a period of 62.5 nm. Another prominent peak at similar length scale occurs at sampling frequency corresponding to a period
of 91.0 nm (which is about 1.5 times 62.5 nm). Similarly, for the section C1, the Fourier analysis revealed a prominent peak at the spatial frequency corresponding to a period of 65.5 nm. Fourier spectra of these sections and a synthetic example with periodically placed emitters are provided in Figs. \ref{fig:actin_period_FFT} and \ref{fig:SynPeriod}, respectively.

Chiu et. al~\cite{chiu1999high} reported the period of the actin double helix to be 37 nm, which corresponds to a period of 74 nm for a single helix. However, actin is known to have structural polymorphs \cite{galkin2010structural} and disorders of sub-helical dimensions \cite{egelman1982f}. Additionally, poly-lysine, used in these samples for stabilizing the actin filaments on the glass chamber, is known to introduce polymeric variations \cite{fowler1982polymorphism}. Thus, we suggest that this periodicity is associated with the periodicity of a single helix of the double helical structure.

\vspace{2mm}
\noindent\textbf{The performance of MUSICAL for non-sparse blinking \textemdash{}}
As noted before, most localization approaches rely on long dark states
of fluorophores (sparse blinking) such that the few bright fluorophores
are optically resolvable. MUSICAL does not have this restriction. To illustrate this, we consider in-vitro sample 2 which exhibits high density of fluorophores, as illustrated in Fig. \ref{fig:density}. While STORM is not suitable for imaging such sample (see Fig. \ref{fig:STORM_sample2}), MUSICAL can image these samples efficiently as seen in the second and fourth rows of Fig. \ref{fig:nonSTORM}.

We further demonstrate the ability of MUSICAL to image dense fluorophore structures through the tubulins high density data of Data-SMLM. The results are shown in Fig. \ref{fig:EPFL}(a). The zoom-in of the cyan marked rectangles in the MUSICAL results show that closely lying microtubules with separation smaller than the optical diffraction limit ($\sim$265 nm for this data) can be resolved consistently using MUSICAL.

\vspace{2mm}
\noindent\textbf{Minimum number of frames required for MUSICAL \textemdash{}}
If blinking is sparse, large number of frames have to be acquired to be able to collect fluorescent signal from each fluorophore. However, since sample 2 does not demonstrate sparse blinking, it is interesting to consider if MUSICAL can be applied to such sample using only a few frames.

Owing to the sliding window of size $N$ pixels, the rank of the matrix on which eigenimages are computed is less than or equal to $\min(N,K)$, where $K$ is the number of frames; see sections \ref{sec:core} and \ref{sec:sliding} of the supplement. The number of pixels in the sliding window $N$ is determined by the PSF of the system and thus fixed. Thus, it is desirable to use $K\ge N$.

We have used a sliding window of size $N_w=7$ pixels, which implies that $N=N_w^2=49$. We applied MUSICAL on the first 49 frames of the data acquired for in-vitro sample 2 under excitation power 205.6 W/cm$^2$. The results shown in the fourth column of Fig. \ref{fig:nonSTORM} clearly illustrate the capability of MUSICAL to obtain super-resolution in non-sparse blinking using very few frames. This also indicates the utility of MUSICAL in the study of dynamic biological processes.

We note that emitters that do not emit significant number of photons in the frames used for MUSICAL will not be imaged by MUSICAL. As a further comment on the desirable condition $K\ge N$, we study the effect of the value of $N_w$ in Fig. \ref{fig:window_size}. As demonstrated, soft window of size larger than $N_w$ in section \ref{sec:soft} of the supplement does not add much value while increasing the minimum number of frames and the computation time. Thus, the sufficient number of frames is limited by the blinking rate and emitter density such that all emitters in the sample can blink in order to represent the entire structure and there are enough fluctuations in the intensity over the frames.

\vspace{2mm}
\noindent\textbf{Excitation power and MUSICAL \textemdash{}} Whereas the image stack for sample 1 was acquired under 205.6 W/cm$^2$, we acquired three image stacks for sample 2, each with different excitation powers, viz. 10.3 W/cm$^2$, 40.2 W/cm$^2$, and 205.6 W/cm$^2$. On one hand, the excitation power directly impacts the achievable resolution through the signal strength. On the other hand, it indirectly influences the reconstruction by affecting the blinking rates of fluorophores, where high excitation powers generally result in longer dark states \cite{dempsey2011evaluation}.

The MUSICAL results for these image stacks are shown in Fig. \ref{fig:nonSTORM}. It is seen that at each excitation power, MUSICAL can reconstruct more details than the mean image. The FWHM of the peaks indicated by arrows in Fig. \ref{fig:nonSTORM}(b5,c5) are 49 nm and 61 nm, respectively. The result implies that the FWHM does not deteriorates significantly when the laser power is decreased from 205.6 W/cm$^2$ to 40.2 W/cm$^2$. It also helps to identify the minimum power at which sufficient resolution is achieved while minimizing photo-bleaching. We think that these observations are in general agreement with \cite{dempsey2011evaluation}. We provide an additional example using sample 3 in the supplement and an empirical
plot of MUSICAL FWHM as a function of excitation power in Fig. \ref{fig:power}. The FWHM and side lobes of MUSICAL as functions of the signal to background ratio are given in Fig. \ref{fig:SynEx3}. Effects such as side lobes and camera's noise, that become prominent at very poor signal to background ratio are discussed in Figs. \ref{fig:highpass} and \ref{fig:system_sen}.

\vspace{2mm}
\noindent{\textbf{Comparison between MUSICAL and SMLM techniques \textemdash}} Comparison of MUSICAL and STORM results for regions B and C of sample 1 are shown in Fig. \ref{fig:main_result}. In addition, we compare the results of MUSICAL with SMLM techniques for the Data-SMLM dataset in Fig. \ref{fig:EPFL}(a-c). Implementation and parameters used for SMLM techniques are given in section \ref{sec:parameters} of the supplement. Overlay of MUSICAL images with STORM or PALM images are also provided for tubulins long sequence and tubulinAF647, where MUSICAL images appear in green and STORM or PALM images appear in magenta. We chose these data for overlay and comparison because they demonstrate sparse blinking, and thus favorable for SMLM. The results show good agreement between MUSICAL and SMLM techniques.

Now we compare MUSICAL and SMLM  for differences, such as noted in Fig. \ref{fig:main_result}. Fig. \ref{fig:main_result}(b4,b5) show the cross-sections of B1-B4 for STORM and MUSICAL, respectively. The cross-sections B1-B4 represent a situation of branching of a filament into two, where B1 shows clearly separated branches, B2 shows a line very close to the junction, B3 is approximately on the junction, and B4 is the unbranched filament. In Fig. \ref{fig:main_result}(b4,b5), it is seen that cross-section at B1 show clearly separated branches at the same locations for both MUSICAL and STORM. However, MUSICAL and STORM results at B2 show quite different results. STORM shows approximately 5 peaks at about 193 nm, 233 nm, 320 nm, 380 nm, and 493 nm, while MUSICAL shows three peaks at 161 nm, 228 nm, and 309 nm. It is indeed difficult to conclude the structure at B2. The results of MUSICAL and STORM for B3 are also interesting. MUSICAL shows two separable but closely located maxima (separated by about 33 nm) with the minimum between them being located at about 250 nm. On the other hand, STORM shows a non-smooth plateau spread from about 220 nm to 320 nm. However, at the unbranched actin filament, at cross-section B4, the results of MUSICAL and STORM match very well. We have included examples of synthetic forks with similar geometry but two different blinking rates in Fig. \ref{fig:SynFork}, from which similar results are inferred.

Another interesting example is shown in Fig. \ref{fig:main_result}(c4,c5), where MUSICAL clearly indicates a branching but STORM does not. In the cross-section plots for C2 (just before branching, Fig. \ref{fig:main_result}(c4)) and C3 (after branching, Fig. \ref{fig:main_result}(c5)), it is seen that STORM gives only one maximum while MUSICAL gives two maxima for C3. Also, the maximum of STORM for C3 is closer to the right side maximum generated by MUSICAL.

These differences are attributed to multiple simultaneously bright
fluorophores in the small volume surrounding the junctions. If a couple of closely placed emitters are blinking in a frame in a very small volume, STORM may conclude the presence of one very bright fluorophore at the center of the small volume instead of the actual distribution of the closely placed emitters; unless a multiple-fluorophore fitting approach is incorporated~\cite{small2014fluorophore}. To illustrate this further, we consider a synthetic example SynSTORM, the details of which are provided in section \ref{sec:SynSTORM} of the supplement and the layout is shown in Fig. \ref{fig:EPFL}(d1,d6). The width of each of the five lines is about 7 nm such that each line mimics an actin filament. The two closest and parallel lines are separated by 50 nm. The top row corresponds to longer dark states ($\tau_{{\rm on}}=0.05{\rm s}$, $\tau_{{\rm off}}=250{\rm s}$) while the bottom row corresponds to shorter dark states ($\tau_{{\rm on}}=0.05{\rm s}$, $\tau_{{\rm off}}=50{\rm s}$), where $\tau_{{\rm on}}$ and $\tau_{{\rm off}}$ represent average time spent by molecules in the bright (on) and dark (off) states respectively.

In addition to the imaging result of STORM shown in Fig. \ref{fig:EPFL}(d4,d9), we also provide localization results of STORM in Fig. \ref{fig:EPFL}(d5,d10). In the case of long dark states, the number of bright fluorophores at a given time is small even at the junction. Thus, in comparison to MUSICAL result (Fig. \ref{fig:EPFL}(d3)), the performance of STORM is superior in this case. On the other hand, in the case of shorter dark states, the number of bright fluorophores in a small volume along the parallel lines is sufficiently high to render STORM ineffective in resolving the lines (Fig. \ref{fig:EPFL}(d9)). MUSICAL can, however, still resolve the lines. Further, the MUSICAL result and the details reconstructed at the lines and junction are similar in both the cases.

\vspace{2mm}
\noindent{\textbf{Comparison with other super-resolution techniques \textemdash}} Here, we compare MUSICAL with other super-resolution techniques, SOFI \cite{dertinger2009fast}, deconSTORM \cite{mukamel2012statistical}, 3B \cite{cox2012bayesian}, and ESI \cite{ESI}. All these methods use statistical analysis of the fluorescence intensity in the image stack instead of single molecule localization. DeconSTORM combines expectation maximization and deconvolution to form a super-resolved image which converges to the maximum likelihood estimate of emitter locations. It can be used for non-sparse blinking as well. We present comparison of these techniques for region A of in-vitro sample 1, data in Data-SMLM, the test data of 3B, synthetic example SynEx2, and synthetic example with Poisson noise SynEx2SBR3 in Fig. \ref{fig:Comparison}(a-e), respectively. Heavy computational requirements and convergence issue of 3B were found to be limiting in generating the comparison results. Thus, instead of using the entire image stacks, we have used lesser number of frames and smaller region for generating the results for all the methods. We have used only 300 frames from sample 1, all 300 frames from the test data of 3B, and only 200 frames from Data-SMLM. SynEx2 and SynEx2SBR3 have 49 frames each. Overlaid images of results of two methods  in magenta and green colors are included in Fig. \ref{fig:Comparison}(a,c) for the ease of comparison. Colored triangles are used to indicate the respective colors used for the methods in the overlaid images. The details of the implementations and control parameters of these methods used for obtaining the results in Fig. \ref{fig:Comparison} are given in section \ref{sec:parameters} of the supplement.

For region A of sample 1 in Fig. \ref{fig:Comparison}(a), MUSICAL and 3B perform similar and provide better result than deconSTORM, ESI, and SOFI. For Data-SMLM, MUSICAL results in sharper images than 3B and deconSTROM, as seen in Fig. \ref{fig:Comparison}(b). In the middle row of Fig. \ref{fig:Comparison}(b), the cyan arrows indicate an instance where MUSICAL is able to resolve two microtubules that are unresolved by 3B and deconSTORM. For the test data of 3B, it is seen that the shape details are better constructed by 3B (Fig. \ref{fig:Comparison}(c11)) than any other method including MUSICAL. However, we noted that the vertex of the polygon reconstructed by 3B highlighted using yellow arrows in Fig. \ref{fig:Comparison}(c8,c10-c12) does not agree with the results of the other methods. We are not sure if it is due to a local minimum in the iterative optimization of 3B or it is an accurate location of the vertex.

Lastly, we discuss the comparison of MUSICAL with other methods for synthetic examples SynEx2 and SynEx2SBR3 shown in Fig. \ref{fig:Comparison}(d,e), respectively. The details of these synthetic examples are presented in section \ref{sec:SynEx2} of the supplement. It is seen that MUSICAL gives the sharpest result for the noise-free image stack, followed by deconSTORM and 3B. However, deconSTORM generates some artifacts inside the rings, which seem to be related to the sidelobes that are emphasised in the noiseless data.

For SynEx2SBR3, deconSTORM, 3B and MUSICAL provide reasonable reconstructions, although 3B reconstructs finer rings than MUSICAL. However, 3B generates point-like artifacts as seen along 45$^{\circ}$ and 135$^{\circ}$ directions. This might be the case because of the very low number of frames. DeconSTORM also generates artifacts in the regions where the rings are close to each other. ESI provides good reconstruction, although visible in the logarithmic scale only. We present a summary and comparison of features of various super-resolution techniques in Table \ref{tab:qualitativeComparison} and section \ref{sec:comparison} in the supplement. SynEx2SBR3 is used for comparison of the computation time also. The time taken by each method is reported below its result in Fig. \ref{fig:Comparison}(e). While MUSICAL takes significantly larger computation time than SOFI and ESI, it compares well with deconSTORM and is orders of magnitudes faster than 3B. We note that no parallelization has been used for reporting the computation time of MUSICAL, although there is a scope for parallelization and thus improvement in the computation time of MUSICAL.

\vspace{2mm}
\noindent{{\textbf{Live-cell experiments \textemdash}} Two sets of live-cell experiments were conducted on Chinese hamster ovary cells (CHO-K). The cells were imaged using TIRF microscope under physiologically relevant conditions. In the first set of experiments, we imaged microtubules labeled by transiently expressed GFP-tubulin. MUSICAL images for this case are presented below. Some additional results are included in the supplement. In the other set, we imaged GFP-labelled actin cytoskeleton in a CHO-K1 cells stably expressing Lifeact-GFP. The images for this case are shown in supplement.

\begin{figure}[ht!]
  \centering
  \includegraphics[width=\linewidth]{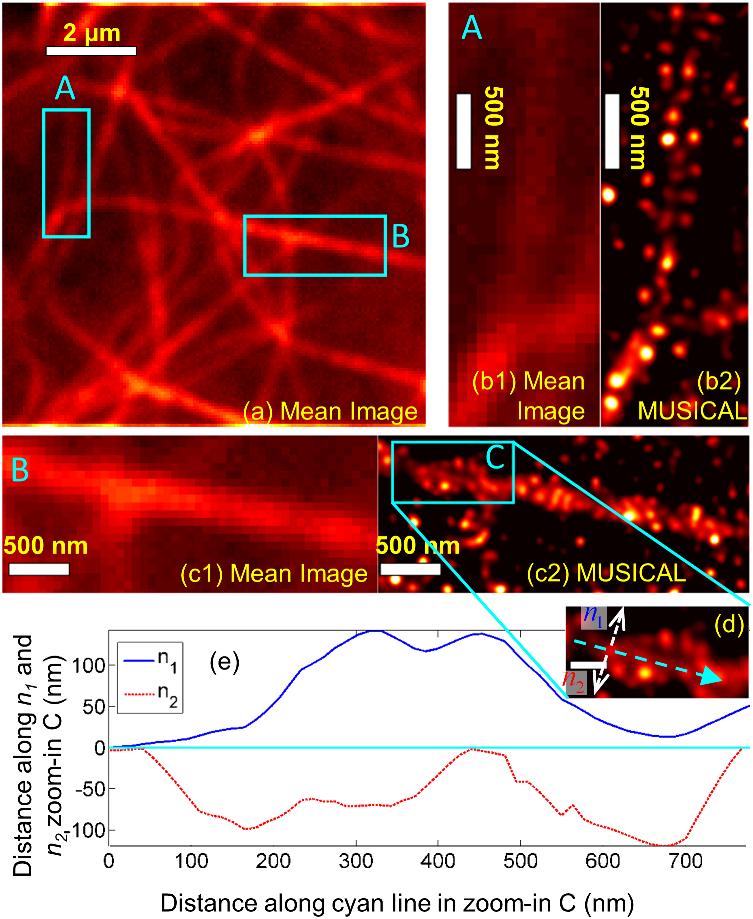}\\
  \caption{Result of MUSICAL for live-cell microtubules sample 1 is shown here. Mean image of 49 frames is shown in (a). Mean image and MUSICAL result for region A identified in (a) are shown in (b1,b2). Mean image and MUSICAL result for region B identified in (a) are shown in (c1,c2). Zoom-in of region C identified in (c2) is shown in (d); scale bar is 200 nm. Subfigure (e) shows the closest maxima from points along the cyan line in (d) in the directions $n_1$ (blue) and $n_2$ (red).}\label{fig:LiveMicrotubules}
\end{figure}

In the experiment with live-cell microtubules, we used regular GFP and standard live cell imaging medium with no additive chemicals to influence blinking. Thus, the cells were in physiologically relevant conditions optimally suited for biological experiments. 488 nm laser excitation at the intensity of 25 W/cm$^{2}$ was used. In live cells, microtubules are highly dynamic structures and there is a continuous exchange of tubulin molecules between the microtubules and cytosol \cite{nogales2001structural}. Therefore, there is a significant population of labelled tubulin molecules freely diffusing in the cytosol in addition to the labelled microtubules. In order to suppress the large background introduced by these freely diffusing molecules, we have heuristically used a lower bound of 0.4 times the MUSICAL intensity that corresponds to 99.9\% of the histogram of intensities. This choice is analogous to the heuristic choice of the upper and lower bounds used in 3B~\cite{cox2012bayesian}. Further, only a part of the population of tubulin molecules in the cell is labelled. Thus, the microtubules are not uniformly labelled and fluorescent in parts only.

MUSICAL result for live-cell microtubules sample 1 is shown in Fig. \ref{fig:LiveMicrotubules}. This result is generated using the first 49 frames, each with 5 ms exposure, which amounts to a total time 245 ms for a MUSICAL image. The zoom-ins shown in Fig. \ref{fig:LiveMicrotubules}(c,d) indicate that MUSICAL provides resolution of less than 100 nm in the live cells too. We note that the overlay of the microtubules as seen in the zoom-in has been observed before \cite{fawcett1993bloom,amos1974arrangement} and not considered particularly significant. We use this structure to illustrate sub-100nm resolution and the reconstruction of details by MUSICAL. This is further validated quantitatively in Fig. \ref{fig:LiveMicrotubules}(e), which plots the distance of the closest maximum along the direction $n_1$ (or $n_2$) from the cyan colored line shown in Fig. \ref{fig:LiveMicrotubules}(d). It shows that the three microtubules can be identified separately even though the distance between them is quite small. The effect and benefit of using lower number of frames in live-cell experiment is demonstrated in Fig. \ref{fig:LiveCellsSample1TimeLapse} of the supplement, which shows results for region B of live-cell microtubules sample 1 with different frame numbers taken from different parts of the whole image stack. Further, we show using the results of live-cell microtubules sample 2 in Fig \ref{fig:LiveCellsSample2} that the time scales of MUSICAL may potentially be reduced to less than 50 ms.

\section*{Discussion}

Although the reported resolution of STORM at 20 nm is significantly better than MUSICAL, MUSICAL demonstrates significantly less stringent requirements on sparsity of blinking, number of frames, excitation power, etc. Further comparison of MUSICAL with other statistical super-resolution techniques, such as SOFI, deconSTORM, 3B, and ESI  indicates competitive and often superior performance of MUSICAL. Here, we mention that MUSICAL performs better than SOFI and ESI in terms of resolution and image quality, however at the cost of increased computational time and using the knowledge of the point spread function. Analysis of sensitivity of MUSICAL to the accurate knowledge of the optical PSF is provided in Fig. \ref{fig:PSF_sensitivity} and section \ref{sec:PSF_sensitivity}. It indicates that MUSICAL is not too sensitive to PSF. Although 3B shows better robustness to noise than MUSICAL, it has significantly higher computational demands unless faster implementation of 3B using cloud computation \cite{hu2013accelerating} is used.

Lastly, to the best of our knowledge, none of the currently known super-resolution algorithms so far has demonstrated the capability to produce super-resolved images of live cells in their natural environment with total acquisition times of the order of 50-250 milliseconds using regular non-sparse blinking dyes. We do note that Huang et. al \cite{huang2013video} have reported super-resolution at video acquisition rate of 32 frames per second, although using photoswitchable proteins or Alexa Fluor 647 in glucose oxidase and catalase based imaging medium, as opposed to the live cell results reported here which use regular GFP and do not use oxygen scavenging and carbon-dioxide deficient imaging medium considered toxic to live-cells \cite{ettinger2014fluorescence}. We believe that MUSICAL is a step forward towards the goal of fast super-resolved imaging of biological phenomena in their natural state.

\textbf{Source codes of MUSICAL and the data generated for this article are available at} \url{https://sites.google.com/site/uthkrishth/musical}

\section*{Methods}

\noindent \textbf{In-vitro actin sample preparation}

We used the protocol suggested in \cite{metcalf2013test} for forming the in-vitro actin samples. Preformed actin filaments from Cytoskeleton Inc., extracted from rabbit skeletal muscle, were used for forming the sample. In one chamber of an 8-chambered cover glass of about 750 ${\rm \mu}$l volume, 200 ${\rm \mu}$l of 0.01\% poly-L-Lysine solution was incubated for 10 minutes. Remaining liquid was drained out using a pipette. 90 ${\rm \mu}$l of general actin buffer (reconstituted as suggested by Cytoskeleton's product datasheet of general actin buffer) was introduced in the chamber. This was followed by adding 10 ${\rm \mu}$l of 10${\rm \mu}$M preformed actin filaments and 10${\rm \mu}$l of Phalloidin Atto-565 solution (stock solution prepared as recommended by the manufacturer, Sigma-Aldrich). The contents of chamber were mixed by gently pipetting up and down. An incubation time of about 45 minutes was allowed. Then the liquid in the chamber was removed by pipetting. For sample 1, the contents of the chamber were washed 3 times using Buffer B (see details of imaging buffer below) by a two pipette system, where 1 pipette let in the buffer and the other pipette pipetted out the solution. Sample 2 was washed 5 times. Sample 3, with results in the supplement, was not washed at all. In sample 2, we introduced tetraspeck beads in the imaging solution and allowed the solution to settle for three hours.

\noindent \textbf{Preparation and introduction of imaging buffer}

We used the imaging buffer composition suggested in \cite{NikonSTORM}. Buffer A composed of 10 mM of TRIS (pH 8.0) and 50 mM NaCl. Buffer B composed of 50 mM TRIS (pH 8.0), 10 mM NaCl, and 10\% Glucose (weight per volume). GLOX solution (1 ml) was formed by vortex mixing a solution of 56 mg of Glucose oxidase, 200 ${\rm \mu}$l of Catalase (17 mg/ml) and 800 ${\rm \mu}$l of Buffer A. MEA solution (1M, 1 ml) was prepared using 77 mg of MEA and 1 ml of 0.25N HCl. For one chamber of an 8-chambered cover glass of about 750 ${\rm \mu}$l volume, 700 ${\rm \mu}$l of imaging buffer was prepared by mixing 7${\rm \mu}$l of GLOX solution, 70 ${\rm \mu}$l of MEA solution, and 620 ${\rm \mu}$l of Buffer B on ice. This imaging buffer was used as the medium for in-vitro actin samples. The chamber was filled completely and covered immediately to avoid replenishing of oxygen in solution. \vspace{2mm}

\noindent \textbf{Live-cell sample preparation}

CHO-K1 cells were obtained from ATCC (Manassas, VA). Lifeact cells (CHO-K1 cells stably expressing Lifeact-GFP) were provided by Prof. Rachel S. Kraut (NTU, Singapore). CHO-K1 and Lifeact cells were cultivated in DMEM medium (Dulbecco's Modified Eagle Medium, Invitrogen; Singapore) supplemented with 1\% penicillin G and streptomycin (PS, PAA, Austria), and 10\% fetal bovine serum (FBS, Invitrogen; Singapore) at 37$^\circ$ C in 5\% (v/v) CO$_{2}$ environment. GFP-tubulin plasmid was a gift from Dr. Pakorn T. Kanchanawong (MBI, NUS, Singapore). Electroporation was used for transfection of the cells, during which, ~90\% confluent cells in a 75 cm$^{2}$ flask were washed twice with 1$\times$PBS, trypsinized with 0.25\% trypsin-0.03\% EDTA solution for approximately 1 minute at 37$^\circ$ C, and then re-suspended in culture medium. Cells were precipitated by centrifugation and re-suspended in small amount of resuspension R buffer (NeonTM Transfection System, Life Technologies, Singapore) and transferred half into one electroporation cuvette (2 mm wide, Bio-Rad; Hercules, CA) for one transfection. Between 300 and 500 ng/ml of the plasmid were added. After electroporation pulse, cells were seeded back to prewashed cover-glass (30 mm in diameter; Lakeside, Monee, IL) in a 35 mm culture dish. Transfected cells grew in the culture medium for 24-36 hours before measurement. All cells were imaged in Phenol red free DMEM medium containing 10\% FBS.
\vspace{2mm}

\noindent \textbf{Imaging system and image acquisition}

The setup consisted of an inverted epi-fluorescence microscope (IX83, Olympus, Japan) equipped with a motorized TIRF illumination combiner (IX3-MITICO, Olympus, Japan) and a scientific complementary metal oxide (sCMOS) camera with 6.5 ${\rm \mu}$m pixels (Orca-Flash4.0, Hamamtsu Photonics, Japan). A 488 nm laser (LAS/488/100/D) or 561 nm laser (LAS/561/100, Olympus, Germany) was connected to the TIRF illumination combiner in which the incidence angle for was adjusted to give 110 nm penetration depth of the evanescent field. The 488 nm and 561 nm lasers were used for imaging live-cell samples or in-vitro actin filaments, respectively. A 100x, NA 1.49 oil immersion objective (UAPON, Olympus, Japan) was used to illuminate the sample and collect the fluorescence image. The fluorescence light then passed through a major dichroic (ZT405/488/561/647rpc, Chroma Technology, Bellows Falls, VT) and a band-pass filter (ZET405/488/561/647m, Chroma Technology, Bellows Falls, VT).

The intensity of the excitation light was determined as the power of the laser light exiting the objective divided by the illuminated area; the power of the laser light was regulated by the proprietary laser control software and by an additional OD 1 neutral density filter. The camera was controlled by Micro-Manager 1.4 \cite{edelstein2014advanced}. For multiple power measurements, the first measurement is done using the lowest power and the power is subsequently increased. The optical point spread function computed as an airy disk has FWHM of about 198 nm for these parameters and emission wavelength 593 nm of the phalloidin Atto-565 dye and about 176 nm for emission wavelength 512 nm of the lifeact GFP.

For in-vitro samples 1 and 2, 10,000 frames were
acquired with exposure time of 5ms (200 frames per second) For in-vitro sample 3, 20,000 frames were acquired with exposure time of 10 ms (100 frames per second).  For live-cell microtubule sample 1, 1,000 frames were acquired with exposure time of 5ms (200 frames per second). For live-cell microtubule sample 2, 49 frames were acquired with exposure time of 1 ms (1,000 frames per second). For live-cell F-actin sample, 100 frames were acquired with exposure time of 1 ms (1,000 frames per second). \vspace{2mm}

\section*{Acknowledgements}

The research was conducted as a part of SMART seed project grant number S900154. The authors gratefully acknowledge the help of Ms. Shuangru Huang with cell culture and transfection, the gift of GFP-tubulin plasmid from Dr. Pakorn T. Kanchanawong, and the gift of Lifeact cells from Prof. Rachel S. Kraut. The authors thank the providers of the online data and source codes used in this paper. The authors acknowledge discussion with Dr. Lu Gan, Dr. Kalpesh Mehta, Dr. Dilip K. Prasad, Dr. Sreelatha Sarangpani, and Miss. Shi Hua Teo. Authors especially thank Dr. Dilip K. Prasad for several technical discussions. The authors declare that they have no competing financial interests. Correspondence and requests for materials should be addressed to Krishna Agarwal (email: uthkrishth@gmail.com).

\bibliographystyle{naturemag}

\clearpage

\newcommand{\ignore}[1]{}

\setcounter{equation}{0}
\setcounter{figure}{0}
\setcounter{table}{0}
\setcounter{page}{1}
\makeatletter
\renewcommand{\thepage}{S.\arabic{page}}
\renewcommand{\theequation}{S\arabic{equation}}
\renewcommand{\thefigure}{S\arabic{figure}}
\renewcommand{\thetable}{S\arabic{table}}
\renewcommand{\@biblabel}[1]{[S#1]}
\renewcommand{\citenumfont}[1]{S#1}
\makeatletter

\twocolumn[\begin{center}
\vspace{10mm}
\textbf{\huge {Mutliple Signal Classification Algorithm (MUSICAL) for super-resolution fluorescence microscopy}}

\vspace{5mm}
\huge {\textbf{ - Supplement}}

\vspace{10mm}

\large Table of contents

\vspace{5mm}
\end{center}]
\makeatother

\renewcommand{\contentsname}{}
\tableofcontents{}

\clearpage
\section{Introduction and outline}

This document is a supplement to the main paper titled as above. The mathematical and algorithmic details of MUSICAL are presented in section \ref{sec:MUSICAL}. Synthetic examples to illustrate various aspects of MUSICAL are presented in section \ref{sec:Synthetic}. Additional
results on in-vivo samples are given in section \ref{sec:invivo}. More details and extra studies on in-vitro samples and analogous synthetic examples are given in section \ref{sec:Results}. Comparison of MUSICAL with other methods
that rely principally on statistical analysis of blinking for achieving super-resolution are
presented in section \ref{sec:comparison}.

\section{MUSICAL - concept and algorithm}\label{sec:MUSICAL}

We first begin with the mathematical model for the temporal image stack formed by blinking fluorophores. Then, the core concept of MUSICAL is presented. The MUSICAL algorithm is then presented and it is followed up by a discussion on the choice of threshold value $\sigma_{0}$ (introduced later).

\subsection{Mathematical background: modeling of the image of blinking fluorophores}

For the ease of reference, we denote points in the image plane as $\vec{r}$ and the sample plane as $\vec{r}'$. For simplicity, we assume that contribution from off-plane emitters can be ignored, such that the sample can be considered two-dimensional. The $x$ and $y$ coordinates (i.e. lateral coordinates) of the points in the image plane are represented as $(x,y)$. Similarly, the lateral coordinates of the points in the sample plane are represented as $(x',y')$. Necessary subscripts may be added as relevant.

Let there be $M$ emitters (individual blinking fluorophores) located at $\vec{r}'_{m},m=1\,\,{\rm {to}}\,\,M$ in the sample plane of a microscopy system. We denote the temporal blinking state of the $m$th emitter as a binary signal $b_{m}(t)$, its emission strength as $e_{m}$, and the emission wavelength as $\lambda$.

Let the image acquisition rate be given by $\Delta t^{-1}$ frames per second, where $\Delta t$ is the image acquisition time per frame. The total emission from the emitter during the $k$th frame is given by
\begin{equation}
s_{m}(k)=e_{m}\int_{(k-1)\Delta t}^{k\Delta t}b_{m}(t)\,{\rm d}t.\label{eq:emission}
\end{equation}

In the imaging system, we denote the centers of the regularly arranged pixels in the image plane as $\vec{r}_{n};n=1\,\,{\rm to}\,\,N$ where $N$ is the number of pixels. The pixel is assumed to be square of width $w$. We interpret the point spread function (PSF) $G(\vec{r},\vec{r}')$ of the imaging system as a mapping which maps the emission from a point $\vec{r}'$ in the sample plane to the intensity at a point $\vec{r}$ in the image plane. The intensity measured at the $n$th pixel is given by
\begin{equation}
{I_{k}}({\vec{r}_{n}})=\int\limits _{\left({{y_{n}}-\frac{w}{2}}\right)}^{\left({{y_{n}}+\frac{w}{2}}\right)}{\int\limits _{\left({{x_{n}}-\frac{w}{2}}\right)}^{\left({{x_{n}}+\frac{w}{2}}\right)}{\sum\limits _{m=1}^{M}{G(\vec{r},{{\vec{r}'}_{m}}){s_{m}}(k){\rm {d}}x{\rm {d}}y}}}\:,\label{eq:int}
\end{equation}

We assume a diffraction limited imaging system in which the dimensions of pixels are significantly smaller than the extent of the main lobe of the PSF of the system. For an imaging
system whose PSF can be approximated as an Airy disk, this implies
that $w\ll1.22\lambda/{\rm NA}$, where ${\rm NA}$ is the numerical
aperture of the imaging system. Then, using the approximation
\begin{equation}
\int\limits _{\left({{y_{n}}-\frac{w}{2}}\right)}^{\left({{y_{n}}+\frac{w}{2}}\right)}{\int\limits _{\left({{x_{n}}-\frac{w}{2}}\right)}^{\left({{x_{n}}+\frac{w}{2}}\right)}{G(\vec{r},{{\vec{r}'}_{m}}){\rm {d}}x{\rm {d}}y}} \approx G(\vec{r}_{n},\vec{r}'_{m})w^{2},
\end{equation}
the intensity at the $n$th pixel can be simplified as
\begin{equation}
I_{k}(\vec{r}_{n})=\sum\limits _{m=1}^{M}w^{2}G(\vec{r}_{n},\vec{r}'_{m})s_{m}(k).\label{eq:pixel}
\end{equation}
We consider a simple example to illustrate the nature of $G(\vec{r}_{n},\vec{r}'_{m})$, where $\vec{r}_{n}$ are the centers of the discrete pixels in the image space but the emitter locations $\vec{r}'_{m}$ are not restricted to discrete grid points. The PSF of most incoherent imaging systems can be represented using an Airy disk pattern
\begin{equation}
G(\vec{r},\vec{r}')={\left({\frac{{{{\rm {J}}_{1}}\left({\frac{2\pi {\rm NA}}{\lambda {\rm M}}\left|{{\vec{r}}-{\rm M}{\vec{r}'}}\right|}\right)}}{{\left|{{\vec{r}}-{\rm M}{\vec{r}'}}\right|}}}\right)^{2}}\:,\label{eq:Airy}
\end{equation}
where ${\rm M}$ is the optical magnification of the imaging system and ${\rm {J}_{1}()}$ represents the Bessel function of 1st order and 1st kind. Let us assume and imaging system with magnification 1 and NA=1.49. Emission wavelength of $\lambda=510$ nm is assumed. Let us consider one-dimensional case where $y=0$ and $y'=0$. In Fig. \ref{fig:G_PSF}, we plot the values of $G(\vec{r}_{n},\vec{r}'_{m})$ for $x_{n}=6.5n$ ${\rm \mu m}$, where integer $n$ varies from $-7$ to 7 and four values of $x'_{m}$, all of which correspond to the same pixel area (namely $n=0)$. Thus, while $G(\vec{r},\vec{r}')$ is a point-to-point mapping, $G(\vec{r}_{n},\vec{r}')$ is a point-to-pixel mapping. MUSICAL uses the point-to-pixel mapping $G(\vec{r}_{n},\vec{r}')$.

\begin{figure}
\centering \includegraphics[width=1\linewidth]{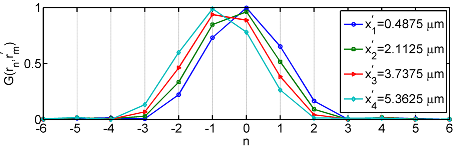}
\vspace{-7mm}
 \caption{Illustration of the effect of imaging approximation on the mapping
vector $G(\vec{r}_{n},\vec{r}'_{m})$.}
\label{fig:G_PSF}
\end{figure}

Intensity measurements in the $k$th frame at all the pixels can be
written as a matrix equation as follows:
\begin{equation}
\bar{I}_{k}={\bf G}\bar{s}_{k}\,\,,\label{eq:Image_frame}
\end{equation}
where,
\begin{equation}
\bar{I}_{k}=\left[\begin{array}{cccc}
I_{k}(\vec{r}_{1}) & I_{k}(\vec{r}_{2}) & \ldots & I_{k}(\vec{r}_{N})\end{array}\right]^{{\rm T}}\,\,,\label{eq:vector_I}
\end{equation}
\begin{equation}
{\bf G}=w^{2}\left[\begin{array}{cccc}
\bar{G}(\vec{r}'_{1}) & \bar{G}(\vec{r}'_{2}) & \ldots & \bar{G}(\vec{r}'_{M})\end{array}\right]\,,\label{eq:matrix_G}
\end{equation}
\begin{equation}
\bar{G}(\vec{r}')=\left[\begin{array}{cccc}
G(\vec{r}_{1},\vec{r}') & {G(\vec{r}_{2},\vec{r}')} & {\ldots} & {G(\vec{r}_{N},\vec{r}')}\end{array}\right]^{{\rm T}}\,,\label{eq:vector_G}
\end{equation}
\begin{equation}
\bar{s}_{k}=\left[\begin{array}{cccc}
{s_{1}(k)} & {s_{2}(k)} & {\ldots} & {s_{M}(k)}\end{array}\right]^{{\rm T}}\,\,,\label{eq:vector_S}
\end{equation}
and the superscript $^{{\rm T}}$ denotes vector or matrix transpose. For the ease of further reference, we refer to $\bar{G}(\vec{r}'_{m})$, i.e. mapping from a location \textit{where an emitter is actually present}, as the emitter mapping vector (EMV). This is to differentiate the EMV $\bar{G}(\vec{r}'_{m})$ from the more general vector $\bar{G}(\vec{r}')$.

Lastly, the intensity measurements from all the frames can be collected
together as a 2-dimensional matrix, in which each row corresponds
to a pixel and each column corresponds to a frame. The complete image
stack which contains images from time frames $k=1\,{\rm to}\,K$ can
be written as a matrix

\begin{equation}
{\bf I}=\left[{\bar{I}_{1}\quad\bar{I}_{2}\quad\ldots\quad\bar{I}_{K}}\right].\label{eq:ImageStack}
\end{equation}

\subsection{The core concept of MUSICAL}

\label{sec:core}

The core concept of MUSICAL revolves around the mathematical range of the matrix ${\bf I}$ and its connection to the actual physical system and sample that created the observed intensities in the matrix ${\bf I}$ (thus called the physical definition of the range of the matrix). We discuss this concept below. \vspace{2mm}

\noindent {\textit{Physical definition of the range of the matrix ${\bf I}$} \textemdash } It is evident from eqs. (\ref{eq:Image_frame},\ref{eq:ImageStack}) that each vector in ${\bf I}$ is a linear combination of the EMVs $\bar{G}(\vec{r}'_{m}),\,m=1\,\,{\rm to}\,\,M$. Thus, the range of the matrix ${\bf I}$ is spanned by the EMVs ${\bf R}:\left\{ \bar{G}(\vec{r}'_{m});m=1\,{\rm to}\,M\right\} $. \vspace{2mm}

\noindent {\textit{Mathematical definition of the range of the matrix ${\bf I}$} \textemdash } Singular value decomposition (SVD) of the matrix ${\bf I}$ gives the spatial basis vectors $\bar{u}_{\sigma_{i}}$ and temporal basis vectors $\bar{v}_{\sigma_{i}}$ with associated singular values $\sigma_{i}$, such that ${\bf I}\,\bar{v}_{\sigma_{i}} = \sigma_i \bar{u}_{\sigma_{i}}$. The range of the matrix ${\bf I}$ is thus given by ${\bf R}:\left\{ {\bar{u}_{\sigma_{i}\ne0}}\right\} $. The vectors $\bar{u}_{\sigma_{i}}$ are referred to as eigenimages and simply written as $\bar{u}_{i}$. \vspace{2mm}

Before proceeding, we consider two cases for the number of emitters $M$. These cases are $M<\min(N,K)$, which corresponds to rank deficiency in the matrix $\mathbf{I}$; and $M\geq\min(N,K)$, which corresponds to the matrix $\mathbf{I}$ being full-ranked. MUSIC, the method that inspired MUSICAL, considered only the rank-deficient case \cite{S-chen2008music,S-agarwal2008applicability}, while MUSICAL deals with both the cases. We note that the null space, which is important to both MUSIC and MUSICAL, is easily determinable in the rank deficient case, whereas the full-ranked case does not have a well defined null space. Thus, we consider the two cases separately first and then use a generalized notation that can be used for both the cases while applying MUSICAL. \vspace{2mm}

\noindent {\textit{Case 1: Rank deficient matrix ${\bf I}$, $M<\min(N,K)$} \textemdash } If the matrix ${\bf I}$ is rank deficient, then there exists a null space ${\bf N}:\left\{ {\bar{u}_{\sigma_{i}=0}}\right\}$ which is orthogonal to ${\bf R}$. The physical definition of the range, discussed above, implies that the EMVs are orthogonal to the vectors in ${\bf N}$, i.e.
\begin{equation}
\bar{G}(\vec{r}'_{m})\cdot{\bar{u}_{\sigma_{i}=0}}=0,\label{eq:MMV_ortho}
\end{equation}
where $\bar{a}\cdot\bar{b}$ denotes the vector dot product of the vectors $\bar{a}$ and $\bar{b}$. Assuming that the fluorophores blink independent of each other, the mapping from $\left\{ \bar{G}(\vec{r}'_{m});m=1\,{\rm to}\,M\right\} $ to $\left\{ {\bar{u}_{\sigma_{i}\ne0}}\right\} $ is one-to-one, i.e. each $\left\{ {\bar{u}_{\sigma_{i}\ne0}}\right\} $ can be represented as a unique linear combination of $\left\{ \bar{G}(\vec{r}'_{m});m=1\,{\rm to}\,M\right\} $ and vice versa. Consequently, the mapping vectors at other location $\vec{r}'\notin\{\vec{r}'_{m},\,\,m=1\,{\rm to}\,M\}$ have a non-zero projection on the null space ${\bf N}$, i.e.
\begin{equation}
\bar{G}(\vec{r}')\cdot{\bar{u}_{\sigma_{i}=0}}\ne0,\,\,{\rm if}\,\,\vec{r}'\notin\{\vec{r}'_{m},\,\,m=1\,{\rm to}\,M\}.\label{eq:non_MMV_ortho}
\end{equation}
Thus, an indicator function can be designed which tests the conditions in eqs. (\ref{eq:MMV_ortho},\ref{eq:non_MMV_ortho}) at different test points $\vec{r}'$ in the sample plane. \vspace{2mm}

\noindent {\textit{Case 2: Full-rank matrix ${\bf I}$}, $M\geq\min(N,K)$ \textemdash } If $M\geq\min(N,K)$, then the rank of the matrix is $\min(N,K)$, none of the singular values is zero, and consequently a null space does not exist. However, the eigenimages $\bar{u}_{i}$ now correspond to different structural details of the arrangement of the fluorophores and their corresponding $\sigma_{i}^{2}$ represent the energy (or strength) of the eigenimage $\bar{u}_{i}$. We can choose first few eigenimages with the large eigenvalues as the representative of the structure and assume that the presence of noise may corrupt the details in the eigenimages with smaller eigenvalues. Quantitatively, we can choose a threshold $\sigma_{0}$ such that $\sigma_{i}<\sigma_{0}$ is considered small and the corresponding eigenimages can be designated as belonging to the null space ${\bf N}:\left\{ {\bar{u}_{\sigma_{i}<\sigma_{0}}}\right\} $ and the space orthogonal to it as the range space ${\bf R}:\left\{ {\bar{u}_{\sigma_{i}\ge\sigma_{0}}}\right\} $. We note that such designation of the range and the null space through the threshold $\sigma_{0}$ is not mathematically rigorous but practically useful.

As an example, if the fluorophores are arranged in a line, then the eigenimage with the largest eigenvalue indicates the structure of a line, irrespective of the density of molecules along the line. Similarly with a circle. Thus, the major structure pattern of simple arrangements can still be captured with very few eigenimages corresponding to the large singular values.

Thus, if a point $\vec{r}'$ belongs to the structural details characterized by the designated range ${\bf R}$, then $\bar{G}(\vec{r}')$ for such point has zero projection on the designated null space ${\bf N}$, and a non-zero projection otherwise. \vspace{2mm}

\noindent {\textit{Notational generalization of the two cases \textemdash }}
The structural arrangement of fluorophores can be represented by the set of eigenimages corresponding to the large singular values, which we loosely call the range and denote as ${\bf R}$. The cardinality \#${\bf R}$ of ${\bf R}$ is defined as the number of eigenimages in ${\bf R}$. In the case $M<\min(N,K)$, \#${\bf R}$ is equal to $M$ and the range is rigorously defined. In the case $M\geq\min(N,K)$, \#${\bf R}$ is less than $\min(N,K)$ and is user designated. In either case, the null space ${\bf N}$ is the space orthogonal to ${\bf R}$.

An indicator function can be designed to test if $\bar{G}(\vec{r}')\cdot\{\bar{u}_{i}\in{\bf N}\}=0$. If this is so, the vector $\bar{G}(\vec{r}')$ belongs to the structure indicated by ${\bf R}$. Thus the design goals of MUSICAL are
\begin{enumerate}
\item to ensure that \#${\bf R}$ can be determined robustly,
\item to design a suitable indicator function which can provide super-resolution, and
\item to mitigate the effect of noise on the indicator function.
\end{enumerate}

\subsection{Multiple Signal Classification Algorithm}

The flowchart of the multiple signal classification algorithm (MUSICAL) is given in Fig. \ref{fig:MUSICAL}. We discuss each functional block of MUSICAL and its contribution in achieving the goals of MUSICAL.

\begin{figure}
\centering \includegraphics[width=1\linewidth]{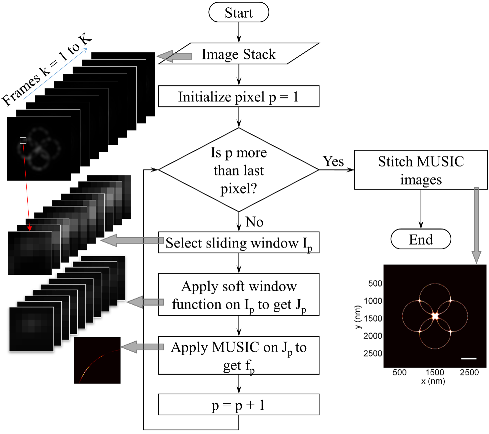}
\vspace{-7mm}
\caption{Flowchart of MUSICAL.}
\label{fig:MUSICAL}
\end{figure}

\subsubsection{Sliding window}\label{sec:sliding}

Instead of considering the whole image stack at once, we consider only small spatial window around a pixel and slide this window across all the pixels. The size of the spatial window is the approximate size of the main lobe of the point spread function (in pixels). Thus, for a PSF resembling an Airy disk, the size of the spatial window can be taken as $N_{w}={\rm floor}\left({\frac{{1.22\lambda}}{{{\rm {NA}}w}}}\right)$. For computational convenience of associating the sliding window with
a pixel, we use the closest odd number,
\begin{equation}
N_{w}=1+2\,{\rm floor}\left({\frac{{0.61\lambda}}{{{\rm {NA}}w}}}\right).\label{eq:N_w}
\end{equation}
The use of floor (i.e., largest integer smaller than the value) implies that we consider the main lobe to an extent close to but smaller than the span between the zeros.

The use of the sliding window contributes towards the goal 1. Choosing a small spatial window instead of the complete image stack helps because there may be several such windows in which the number of fluorophores is small or zero and thus \#${\bf R}$ is easily determinable. Even for the other windows, unless the number of particles in a window is large and the distribution is random or complicated, few eigenimages can represent the distribution of the fluorophores in the window. Thus, \#${\bf R}$ can be reasonably determined. Thus, the use of window directly helps in ensuring the applicability of the aforementioned core concept for most scenarios.

The use of the sliding window also helps towards goal 3. Since the spatial window spans only the main lobe of the PSF, the noise from the pixels beyond the window does not impact the indicator function computed using this window. Moreover, since the fluorophores in center pixel of this window do not contribute significantly to the pixels beyond the main lobe of the PSF, no significant information is lost in the context of the test points inside the center pixel.

For later reference, the window is slid one pixel at a time and the center pixel of the window is denoted as $p$. The image stack corresponding to the window centered at $p$ is denoted as ${\bf I}_{p}$. In the context of section \ref{sec:core}, ${\bf I}$ can be replaced by ${\bf I}_{p}$ and related quantities are calculated with reference to ${\bf I}_{p}$ as the image stack.

\subsubsection{Soft window function}

Due to the use of sliding window, the indicator function is the most reliable for the test points in the center pixel. For the pixels at the edge of the window, the effect of truncation due to the finite size of the window may be considered abrupt. Thus, we apply a soft-window function, which scales the weights of the pixels according to their proximity to the center pixel. This is elegantly achieved in the mathematical framework of section \ref{sec:core}, as discussed next. Let the spatial distribution of the soft window function be given as
\begin{equation}
\bar{h}=\left[{h(\vec{r}_{1})\quad h(\vec{r}_{2})\quad\ldots\quad h(\vec{r}_{N})}\right],\label{eq:soft_fn}
\end{equation}
and the diagonal soft window transformation matrix be given by ${\bf H}={\rm diag}(\bar{h})$. We obtain the soft windowed image stack ${\bf J}_{p}$ as
\begin{equation}
{\bf J}_{p}={\bf H}{\bf I}_{p}.\label{eq:soft_data}
\end{equation}
Further, all mapping vectors are correspondingly transformed as
\begin{equation}
\bar{G}'(\vec{r}')={\bf H}\bar{G}(\vec{r}').\label{eq:soft_map}
\end{equation}
After this step, in the context of section \ref{sec:core}, ${\bf I}$ and $\bar{G}(\vec{r}')$ can be replaced by ${\bf J}_{p}$ and $\bar{G}'(\vec{r}')$, respectively.

We have used 2-dimensional rotationally symmetric Gaussian function, which is centered at the center pixel $\vec{r}_{p}$ and has a spread (standard deviation) $\rho=w\frac{N_{w}-1}{2}$
\begin{equation}
h(\vec{r})=\exp{\frac{\|\vec{r}-\vec{r}_{p}\|^{2}}{2\rho^{2}}}.\label{eq:soft_gauss}
\end{equation}

\subsubsection{Multiple signal classification}

SVD of ${\bf J}_{p}$ yields the eigenimages $\bar{u}_{\sigma_{i}}$ and the corresponding singular values $\sigma_{i}$. In the original form of multiple signal classification (MUSIC), the following indicator function was computed for the test points $\vec{r}'_{{\rm test}}$ in the sample plane
\begin{equation}
f(\vec{r}'_{{\rm test}})=\frac{1}{\sqrt{\sum_{\sigma_{i}<\sigma_{0}}{{\left\Vert {\bar{G}(\vec{r}_{{\rm test}})\cdot\bar{u}_{i}}\right\Vert }^{2}}}}.\label{eq:f_ori}
\end{equation}

The indicator function $f_{(}\vec{r}'_{{\rm test}})$ demonstrates
the following behavior:
\begin{equation}
f({\vec r'_{{\rm{test}}}}) = \left\{ {\begin{array}{*{20}{c}}
\infty &{{\rm{if}}{\mkern 1mu} {\kern 1pt} {\kern 1pt} \;{\kern 1pt} {{\vec r'}_{{\rm{test}}}} \in {\bf{L}}}\\
{{\rm{finite}}}&{{\rm{otherwise}}.}
\end{array}} \right.
\label{eq:classification}
\end{equation}
where ${\bf L}$ represents the fluorophore structures represented by vectors in ${\bf R}$. As a consequence, $f(\vec{r}_{{\rm test}})$ is unbounded individually for two closely placed emitters but bounded as test points between them. This allows two closely placed emitters to be distinguished. We note that even in ideal noiseless measurements, the numerical precision of the computation machine implies that the values are never infinite in practice but limited by the computation precision.

We have modified the indicator function of MUSIC to achieve goal 2, contribute towards goal 3, and enable direct stitching of the MUSIC images of different sliding windows. For the modification, we define the projection distances $d_{{\rm PN}}(\vec{r}_{{\rm test}})$ and $d_{{\rm PR}}(\vec{r}_{{\rm test}})$ as follows
\begin{equation}
d_{{\rm PN}}(\vec{r}'_{{\rm test}})=\sqrt{\sum_{\sigma_{i}<\sigma_{0}}{{\left\Vert {\bar{G}(\vec{r}'_{{\rm test}})\cdot\bar{u}_{i}}\right\Vert }^{2}}},
\end{equation}
\begin{equation}
d_{{\rm PR}}(\vec{r}'_{{\rm test}})=\sqrt{\sum_{\sigma_{i}\geq\sigma_{0}}{{\left\Vert {\bar{G}(\vec{r}'_{{\rm test}})\cdot\bar{u}_{i}}\right\Vert }^{2}}},\label{eq:d_PS}
\end{equation}
where $d_{{\rm PN}}(\vec{r}'_{{\rm test}})$ and $d_{{\rm PR}}(\vec{r}'_{{\rm test}})$ are the projection of $\bar{G}(\vec{r}'_{{\rm test}})$ on the null space ${\bf N}$ and the range ${\bf R}$ respectively. Then, we define the indicator function for MUSIC as
\begin{equation}
f(\vec{r}'_{{\rm test}})=\Bigg({\frac{{d_{{\rm PR}}(\vec{r}'_{{\rm test}})}}{{d_{{\rm PN}}(\vec{r}'_{{\rm test}})}}}\Bigg)^{\alpha},\label{eq:f_mod}
\end{equation}
The behavior of the modified indicator function is similar to eq. (\ref{eq:classification}). It has the following additional salient properties
\begin{itemize}[leftmargin=*]
\item It incorporates the information of all the eigenimages (both ${\bf R}$ and ${\bf N}$) whereas the information of ${\bf R}$ is completely ignored in eq. (\ref{eq:f_ori}).
\item Through the use of $d_{{\rm PR}}(\vec{r}'_{{\rm test}})$, MUSIC image of each sliding window is automatically scaled with reference to the complete image stack and thus, the MUSIC images positioned in the relevant pixel can simply be added for stitching all the MUSIC images and obtaining the MUSICAL result.
\item The value of $\alpha$, typically chosen more than 1, determines the spread of $f(\vec{r}'_{{\rm test}})$ for a single fluorophore. With all the parameters remaining the same, increasing the value of $\alpha$ makes the spread narrower and consequently the resolution better. Thus, in the presence of significant amount of noise, resolution of MUSICAL may be improved by simply choosing larger $\alpha$. On the other hand, it increases the dynamic range of the MUSIC image. Unless mentioned, we have used $\alpha=4$ for generating the MUSICAL results.
\item The modified indicator function is a dimensionless quantity.
\end{itemize}

\subsubsection{Stitching MUSIC images}

Theoretically, the test points $\vec{r}'_{{\rm test}}$ can be any points in the sample region corresponding to the sliding window. But, for the convenience of forming the final image by stitching the results of all the sliding windows, we decimate every pixel into a sub-pixel grid with even number of grid points along each direction and take the centers of the sub-pixels as the test points. Thus, the coordinates of the test points $\vec{r}'_{{\rm test}}$ in a pixel are given as $(x'+(0.5+l_{x})/L_{x},(y'+(0.5+l_{y}))/L_{y}$, where $L_{x},L_{y}$ are the even numbers of sub-pixels per pixel along the $x$ and $y$ directions, $l_{x},l_{y}$ are integers with $|l_{x}|\le L_{x}/2,|l_{y}|\le L_{y}/2$, and $x',y'$ are the conjugate points of the pixel $(x,y)$ in the sample plane.

Suppose we wish to compute the MUSICAL result for the $q$th pixel. When the sliding windows are centered at the pixels with coordinates
\begin{equation}
  x_p=x_q+w n_{x};\quad y_p=y_q+w n_{y}
\end{equation}
where $|n_{x}|,|n_{y}|\leq(N_{w}-1)/2$, then $q$th pixel is a part of these sliding windows and thus MUSIC's indicator function is computed for the test points in the $q$th pixel. Let us denote the indicator function computed for a sliding window with center pixel $(x_p,y_p)$ by $f_{x_p,y_p}(\vec{r}'_{{\rm test}})$. Then, the MUSICAL result at $\vec{r}'_{{\rm test}}$ in the $q$th pixel can be obtained by
\begin{equation}
F\left({{{\vec{r}'}_{{\rm test}}}\;\in \, q{\rm th}\,\, {\rm pixel} }\right)=\sum\limits _{{n_{x}}}{\sum\limits _{{n_{y}}}{{f_{x_p,y_p}}\left({{\vec{r}'}_{{\rm test}}}\right)}}\label{eq:stitch}
\end{equation}

\subsection{MUSICAL and effect of sliding window}\label{sec:soft}
As an example of the effect of the size of window and
the soft window function, we show the result of MUSICAL
for a small portion from region B of in-vitro sample 1 using
different values of $N_w$ in Fig. \ref{fig:window_size}, with and without the soft
window function. According to eq. (\ref{eq:N_w}), the value of Nw for
in-vitro sample 1 is 7. However, the result for $N_w = 7$ without
the soft window function clearly illustrates grid-like artifacts
related to truncation. Further, the images obtained without the
soft window function show incorrect images since they do not
weigh the center pixel at which the PSF is most reliable more
than the other pixels in a window. On the other hand, when the
soft window function is used, the grid-like artifacts are absent
even with $N_w = 7$, Further, appropriate weighing of the pixels
in the window results in more accurate imaging result. Further,
with the use of the soft window function, values of Nw larger
than recommended in eq. (\ref{eq:N_w}) do not change the MUSICAL
result despite using more computational resources.

\begin{figure}
  \centering
  \includegraphics[width=\linewidth]{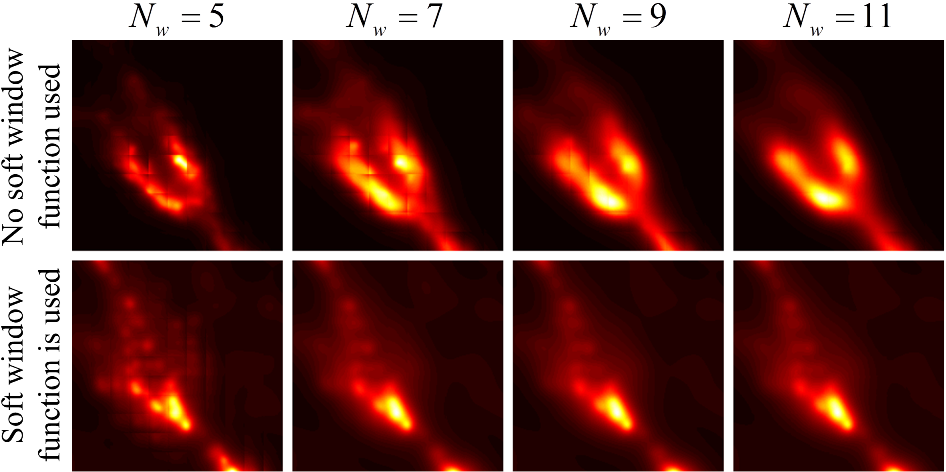}\\
  \caption{The effect of the window size and the soft window function is demonstrated here using a small portion of region A of in-vitro sample 1.}\label{fig:window_size}
\end{figure}

\subsection{MUSIC in the presence of noise and the choice of $\sigma_{0}$}

\label{sec:sigma_0}

The presence of noise affects both the range and the null spaces of the measurement matrix ${\bf I}$ (or the sliding window ${\bf J}_{p}$). As a result of noise, not only the singular values $\sigma_{i}$, but the eigenimages $\bar{u}_{i}$ also change to a certain extent. The most direct effect of noise is that the singular values may no longer be zero even in the rank-deficient case. Yet, the null space can be determined by suitable selection of the threshold $\sigma_{0}$. The indirect impact of noise is that $d_{{\rm PN}}(\vec{r}'_{{\rm test}})\neq0$ for $\vec{r}'_{{\rm test}}\in{\bf R}$. Thus, the indicator function $f(\vec{r}_{{\rm test}})$ does not have a clear classification as indicated in eq. (\ref{eq:classification}). It rather has a continuous spread as we move away from a point $\vec{r}'_{{\rm test}}\in{\bf R}$. In this situation, higher value of $\alpha$ helps in reducing the spread.

We provide two rules of thumb to select the value of the threshold
$\sigma_{0}$. If the signal to noise ratio SNR = $\|{\bf I}\|/\|{\bf N}\|$ is known, then $\sigma_{0}$ is chosen as ${\sigma_{\max}}/{\sigma_{0}}={\rm SNR}$, where $\sigma_{\max}$ is the maximum singular value observed for the singular values computed for all the sliding windows. If the SNR is not known, then the value of $\sigma_{0}$ is chosen to be slightly less than the value where a knee feature is observed in the logarithmic plot of
singular values of all the sliding windows. Here, knee feature is characterized by a point before which singular values show a fast decaying characteristics and after which the singular values decay slowly.

\subsection{Discussion on the role of the parameter $\alpha$}\label{sec:alpha}
Here, we discuss further on the role of the parameter $\alpha$
in MUSICAL. Irrespective of the definition of the indicator
function and the value of $\alpha$, resolution between two features is
theoretically achieved in the case of noise-free measurements
if the indicator function is finite at a test point between the
features because the contrast is perfect according to eq. (\ref{eq:classification}).
However, practically the indicator function is not infinite at
the location of the emitters. Further, it is not necessary that
the actual emitter location is also a test point. Nevertheless,
a dip in the value of the indicator at a point between two
features indicates non-zero contrast and thus computationally
resolved features. However, for visual resolution, sufficient
contrast is essential. Since the indicator function of MUSICAL
is inherently non-linear through the presence of $d_{\rm PN}$ in the
denominator, $\alpha$ only tweaks the amount of non-linearity,
consequently influencing three aspects of MUSICAL result.
The first consequence is obvious, it obviously controls the
contrast of the MUSICAL image. The second consequence is
somewhat subtle; the value of $\alpha$ determines the order of nonlinearity
in the stitching of the MUSICAL results of sliding
windows through eq. (\ref{eq:stitch}). The higher the order, the lower is
the contribution of the off-center sliding windows to the final
MUSICAL image. A small value of $\alpha$, such as 1 or 2, implies
larger contribution from the off-center sliding windows which
are more susceptible to noise. A large value of $\alpha$ implies
less contribution from the off-center sliding windows, which
do contain some information about the test points. The third
consequence is that a high value of $\alpha$ non-linearly scales the
less prominent features and makes them visually less apparent.
The features may be less prominent due to lower number of
emissions or less fluctuations. On the other hand, small value
of $\alpha$ emphasises the non-zero and non-uniform background,
such as expected in the MUSIC's indicator function due its
characteristic in eq. (\ref{eq:classification}).
We heuristically found that $\alpha = 4$
provides a good trade-off for all the three aspects in general.

\begin{figure}
  \centering
  \includegraphics[width=\linewidth]{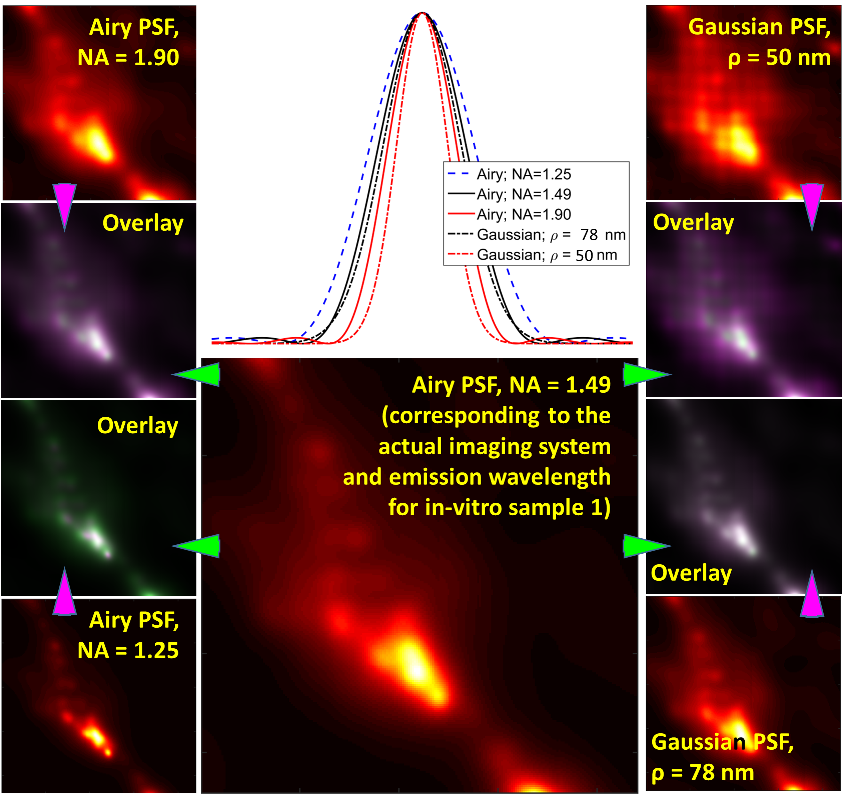}\\
  \caption{An example of sensitivity of MUSICAL to PSF is shown here.
Images on the left side correspond to Airy PSFs with widths not matching
with the actual PSF while the right side corresponds to images with Gaussian
approximations of PSFs.}\label{fig:PSF_sensitivity}
\end{figure}

\subsection{MUSICAL and sensitivity to the optical point spread function}\label{sec:PSF_sensitivity}
We have used the Airy disk function computed using eq.
(\ref{eq:Airy}) as the point spread function throughout the paper. Thus, it
is assumed that the emission wavelength, numerical aperture
(NA), magnification, and pixel width are known. Among these,
only the emission wavelength may not be known accurately.
However, error of a few nm in emission wavelength does not
cause significant error. In case that the PSF is characterized
or calibrated experimentally or does not match the actual PSF
of the system, it is of interest to investigate the sensitivity of
MUSICAL to the incorrect PSF. Here, we present an example
in Fig. \ref{fig:PSF_sensitivity} using a small portion of region B of in-vitro sample 1. We first consider PSFs having the shape of Airy disk but
the width being different from the actual PSF, such as due to a
mismatch between the actual and the estimated NA. When a wider
PSF is used (lower NA), MUSICAL image appears sharper and more punctuated as
seen in bottom left images in Fig. \ref{fig:PSF_sensitivity} and vice versa for a
narrower PSF (top left images in Fig. \ref{fig:PSF_sensitivity}). Next, we consider
the case where the width of the PSF is approximately the same
as the actual PSF, but the shape is approximated as a Gaussian
function instead of an Airy disk. In this case, the MUSICAL
result is almost the same for the actual and Gaussian approximated PSFs.
However, if the Gaussian approximated PSF is too narrow, then grid-like
artifacts appear in the image. In conclusion, MUSICAL is not
very sensitive to the shape and width of PSF. Nevertheless, it
is significantly more sensitive to the PSF than most single
molecule localization techniques and 3B. In our observation,
deconSTORM also demonstrates similar sensitivity to PSF
because of the sensitivity of deconvolution to the estimate of
PSF.

\section{Synthetic examples}\label{sec:Synthetic}
This section provides synthetic examples to illustrate important algorithmic features of MUSICAL introduced in section \ref{sec:MUSICAL}. All the examples assume an emission wavelength $\lambda=510{\rm nm}$, a TIRF system of NA =1.49, magnification 100X, and sCMOS camera with pixel size 65 nm. The simulation of blinking for each of the examples is also provided in this section.

\begin{figure}[t]
\centering \includegraphics[width=1\linewidth]{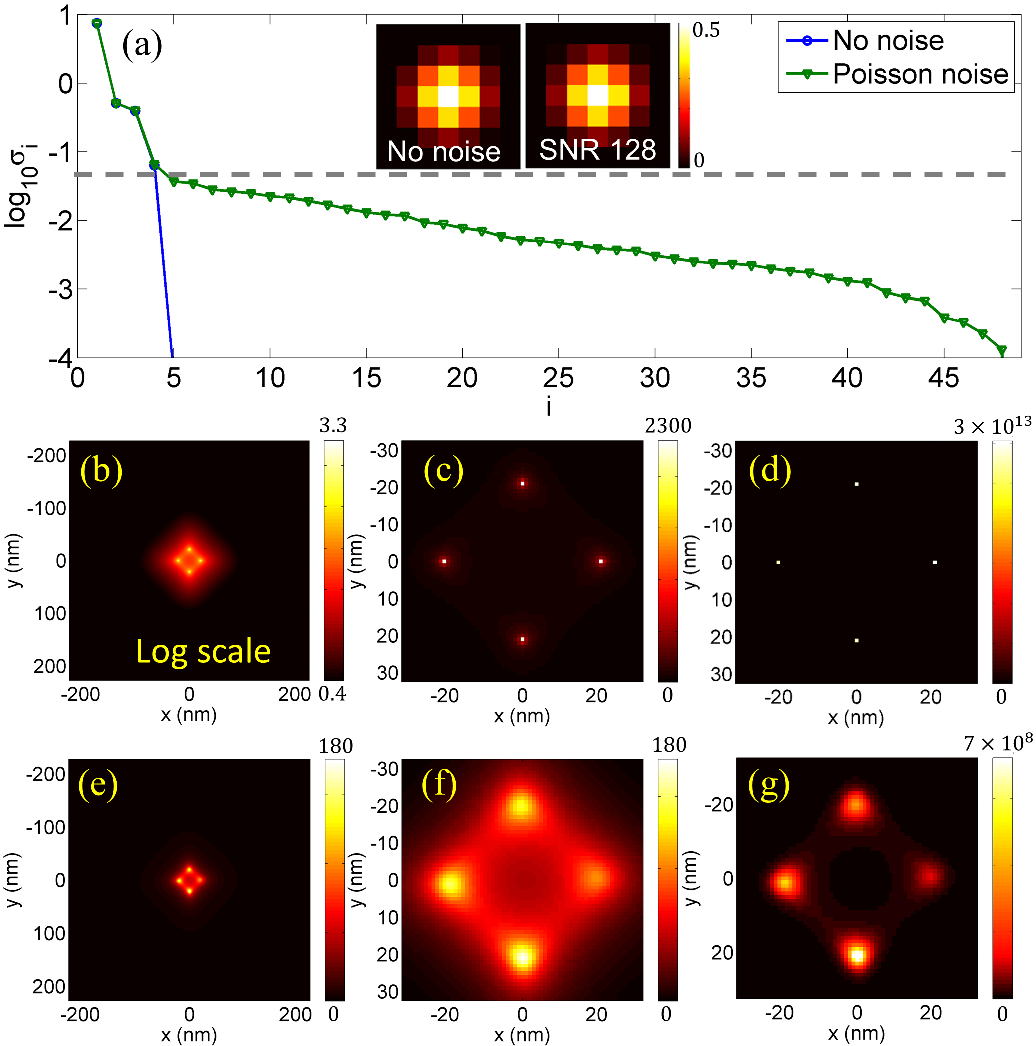}
\vspace{-7mm}
\caption{Synthetic example 1 illustrates the effect of noise on the modified MUSIC indicator function and the effect of the value of $\alpha$. (a) Plot of singular values. Insets show the mean image of the image stacks with and without noise. Middle row corresponds to data without noise while Bottom row corresponds to data with shot noise such that signal to noise ratio is 128. (b,c,e,f) correspond to $\alpha=1$, where (c,f) are zoom-in of the central region of (b,e). (d,g) correspond to $\alpha=4$ and show the same region as shown in (c,f). (b) is shown in logarithmic scale for ease of visualization. The color bar of (g) has been made slightly more red for the ease of visualization of the right most emitter.}
\label{fig:SynEx1}
\end{figure}

\subsection{Example with 4 emitters \textemdash SynEx1}

We illustrate the effect of noise and the value of $\alpha$ on the modified indicator function using synthetic example SynEx1. This example consists of four emitters arranged at the corners of a square of size 30 nm whose corners lie on the $x$ and $y$ axes. For this example, we assume that the image is captured on $7\times7$ pixel array only, which corresponds to $N_{w}$ in eq. (\ref{eq:N_w}). We consider $K=49$ frames. There is only one window in this case and thus only MUSIC with modified indicator function is applied here.

It is seen in Fig. \ref{fig:SynEx1}(a) that in the absence of noise, only four singular values are non-zero while the remaining singular values drop to zero or numerical noise of the computer. They are not shown for convenience of visualization. Thus, the null space is clearly defined in this case. The modified indicator function with $\alpha=1$ computed for a grid of test points is plotted in Fig. \ref{fig:SynEx1}(b) and the zoom-in of the center pixel is shown in Fig. \ref{fig:SynEx1}(c). The result clearly shows high values of the indicator function at the test points closest to the emitter locations. Lastly, the use of $\alpha=4$ further reduces the background, as seen in Fig. \ref{fig:SynEx1}(d).

In the presence of noise, all the singular values are non-zero, as seen in Fig. \ref{fig:SynEx1}(a)). Thus, the null space cannot be clearly determined. Nevertheless, a knee can be observed after 4 singular values (indicated using gray dashed lines in Fig. \ref{fig:SynEx1}(a)). Thus, using $\log_{10}\sigma_{0}=-1.2$, where the knee is approximately noted, the modified indicator functions are plotted using $\alpha=1$ in Fig. \ref{fig:SynEx1}(e,f) and using $\alpha=4$ in Fig. \ref{fig:SynEx1}(g). The indicator function shows a continuous spread in the presence of noise and the spread is significantly smaller for $\alpha=4$.

We also show eigenimages and the corresponding eigenvalues
for SynEx1 as an illustration in Fig. \ref{fig:EigenImages}. Only the first
six eigenimages are used for convenience. It is seen that the
eigenimages with non-zero eigenvalues represent information
about the structure and the eigenimages with zero eigenvalues
represent noise patterns. Other eigenimages also correspond
to zero eigenvalues and represent random noise patterns. If
the emitters were not blinking, all the images in the image
stack would be the same and thus only one eigenvalue would
be non-zero and would represent simply the image repeated in
the entire image stack. On the other hand, if the emitters would
be emitting in synchronism with each other, the image stack
could be represented by only two images, one in which all the
emitters are emitting and the other being the dark image. Even
in this case, only one eigenvalue would be non-zero and the
corresponding eigenimage would be the image corresponding
to emissions from the emitters. It is the phenomenon of
blinking that introduces diversity in the images and more than
one non-zero eigenvalues, each of which represents a different
pattern from the image stack.

Fig. \ref{fig:EigenImages} also presents the projections of the PSFs of test
points on the eigenimages. It is seen that the projection
at the actual emitter locations is quite high for the first
eigenimage. In other eigenimages with non-zero eigenvalues,
the projections at the actual emitter locations are smaller than
the projection for the first eigenimage but not equal to zero.
The projections at the actual emitter locations on the fifth and
sixth eigenimages are zero. These properties are more obvious
in the log scale of the projections shown in the bottom row of
Fig. S6. The projections of the PSF are similarly zero at the
emitter locations for all the eigenimages with zero eigenvalues.
At other locations, the projections on the eigenimages with
zero eigenvalues is not consistently zero. These properties of
the projections on the eigenimages in the null space are the
critical feature of MUSIC and MUSICAL and are exploited
in their indicator functions.

In comparison to conventional deconvolution which may
be considered as manipulation in the range space, MUSICAL
does not perform deconvolution in the conventional sense since
it uses splitting of the eigenimages into range and null spaces
and relies more on the projection upon the null space rather
than the projection on the range.

\begin{figure*}
  \centering
  \includegraphics[width=0.75\linewidth]{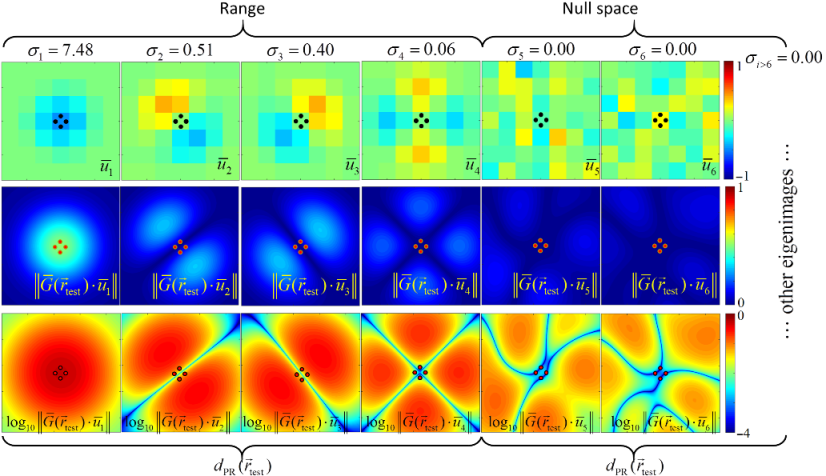}\\
  \caption{The first six eigenimages (top row), the projections of the PSFs of the test points on the eigenimages for SynEx1 (middle row), and the logarithmic
maps of the projections (bottom row) are shown here. Actual emitter locations are shown using black dots in the top row and red dots in the middle and the
bottom rows.}\label{fig:EigenImages}
\end{figure*}

\begin{figure}
\centering \includegraphics[width=1\linewidth]{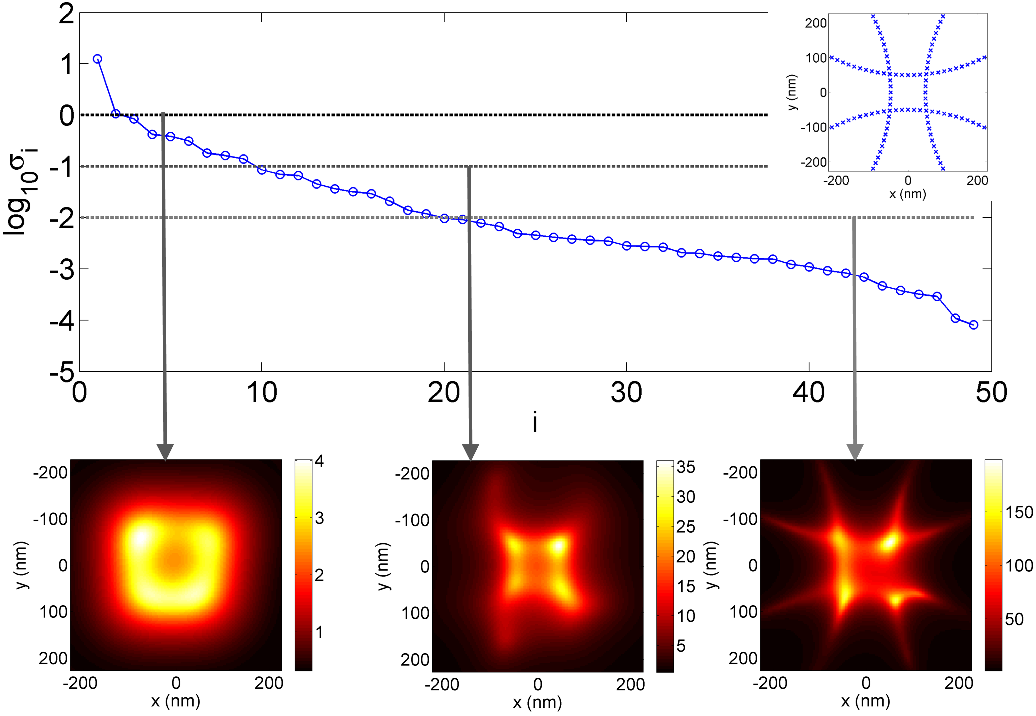}
\vspace{-7mm}
\caption{Synthetic Example 2 in the absence of noise is used to illustrate here the condition $M>\min(N,K)$. Specifically, the sliding window at the center pixel of the image stack is considered. Plot of singular values is shown at the top. Inset shows the emitter distribution. Different horizontal lines correspond to different values of $\sigma_{0}$ and arrows from them point to the corresponding modified indicator function plots ($\alpha=1$ is used).}
\label{fig:SynEx2NoiseLess}
\end{figure}

\subsection{Example with many emitters along four circles \textemdash SynEx2}\label{sec:SynEx2}

We use another synthetic example SynEx2 to illustrate MUSICAL imaging and the effect of the threshold $\sigma_{0}$ for the case $M \ge \min(N,K)$. The example consists of four circles, each of radius 500 nm. Each circle has 180 emitters uniformly located along the circumference of the circle. The centers of the circles are located at the four corners of a square such that the distance between the diagonal corners is 1100 nm and the corners lie on the axes. All the other parameters of the measurement system are the same as SynEx1, except that each image in the temporal image stack is of size $51\times51$ pixels. This example is later used in section \ref{sec:comparison}
as well.

\subsubsection{The role of $\sigma_{0}$}

We study the effect of choosing different values of $\sigma_{0}$ here. We consider the central region of the distribution, i.e. the sliding window for the center most pixel in the image stack. This window of $7\times7$ pixels has 108 emitters. The singular values for this window are shown in Fig. \ref{fig:SynEx2NoiseLess}. It is seen that none of the singular values are zero, indicating that  the null space is not rigorously defined. However, the decay of singular values indicate that we may choose $\sigma_{0}$ to allow different definitions of the null space in the computation of the modified indicator function. For a high value of $\sigma_{0}$, the structure shown in bottom-left of Fig. \ref{fig:SynEx2NoiseLess} can be identified using the modified indicator function. For subsequently smaller value of $\sigma_{0}$, the modified indicator function shows increasingly more details.

\begin{figure}[t]
\centering \includegraphics[width=1\linewidth]{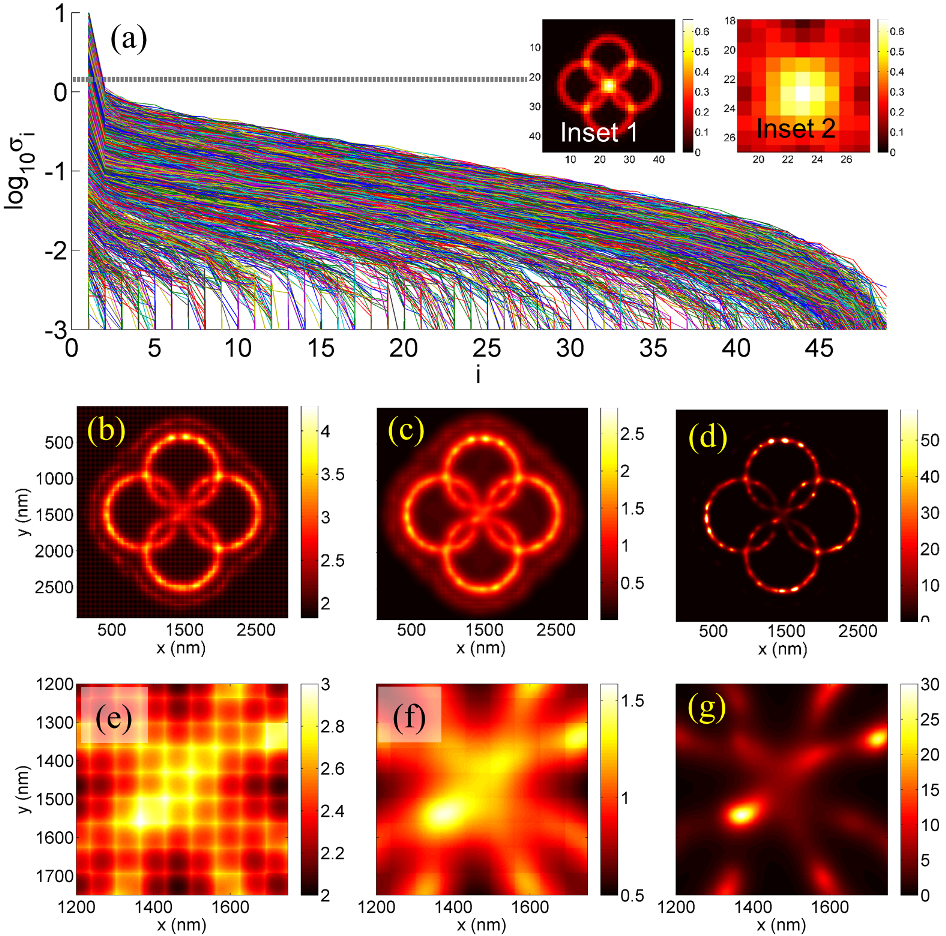}
\vspace{-7mm}
\caption{Synthetic Example 2 with noise shows the effect of the indicator function on smoothing and background values. (a) shows the plot of singular values for all the sliding windows. The gray line indicates $\sigma_{0}$. Further, inset1 and inset 2 show the mean image of the image stack (the MUSICAL results for which are shown in (b-d)) and the central portion (the MUSICAL results for which are shown in (e-g)), respectively. (b-d) show MUSICAL results obtained using MUSIC's original indicator function (b), modified indicator function of MUSICAL with $\alpha=1$ (c), and modified indicator function of MUSICAL with $\alpha=4$ (d). (e-g) are the zoom-ins of the central portions of (b-d) respectively.}
\label{fig:SynEx2}
\end{figure}

\subsubsection{Automatic stitching due to modified indicator function}

Next, we use SynEx2 to demonstrate the automatic stitching ability of the modified MUSICAL indicator function, as opposed to MUSIC's original indicator function. We consider the entire image stack with Poisson noise added to the data such that the signal to noise ratio (SNR) of the noisy image stack is 8. The results of MUSICAL is shown in Fig. \ref{fig:SynEx2}. We compute $\sigma_{0}$ assuming that the SNR is known. The singular values and $\log_{10}\sigma_{0}$ (gray dashed line) are plotted in Fig. \ref{fig:SynEx2}(a). The result of the original indicator function of MUSIC, the modified indicator function of MUSICAL with $\alpha=1$, and the modified indicator function of MUSICAL with $\alpha=4$ are shown in Fig. \ref{fig:SynEx2}(b-d) respectively. Their corresponding zoom-ins of the central region are shown in Fig. \ref{fig:SynEx2}(e-g), respectively.

The artifacts due to the pixel boundaries are clearly visible in the result of the original indicator function, shown in Fig. \ref{fig:SynEx2}(e). These artifacts are almost removed with the modified indicator function with $\alpha=1$ as seen in Fig. \ref{fig:SynEx2}(f). Further, using a higher value of $\alpha$, here $\alpha=4$, the artifacts due to pixel boundaries are completely removed. We have observed that in general, $\alpha\geq2$ does not have the pixel boundary artifacts if the value of $\sigma_{0}$ is close to the optimal value. The halo effect seen prominently in Fig. 5(b,c) is due to the side lobes and is explained later in section \ref{sec:highpass}.

\subsection{Full width at half maximum of MUSICAL}\label{sec:FWHM}
We use synthetic example SynEx3 to quantify the full width at half maximum (FWHM) of MUSICAL at different signal to background ratios, where the background is roughly the offset value of the imaging system. SynEx3 has 500 emitters arranged along a line of 200 nm. The image stack has $K=1000$ frames. The blinking and noise simulations are explained in sections \ref{sec:blinking} and \ref{sec:noise}. Before applying MUSICAL, an offset value of 80 is subtracted from image stack. The value of $\alpha=4$ is used for MUSICAL here onwards. The FWHM is plotted as a function of signal to background ratio (SBR) in Fig. \ref{fig:SynEx3}(a). The minimum value of FWHM in Fig. \ref{fig:SynEx3}(a) is 27.11 nm for SBR 10 and the maximum value is 158.2 nm for SBR 1.2.

\begin{figure}
  \centering
  \includegraphics[width=\linewidth]{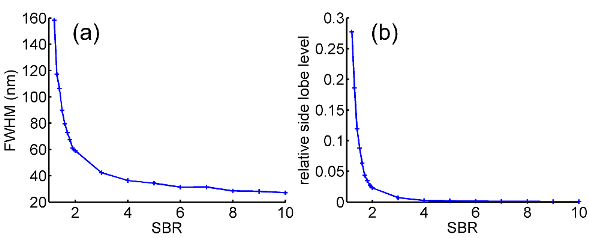}
  \vspace{-7mm}
  \caption{FWHM (a) and side lobes(b) of MUSICAL result for SynEx3 as a function of the signal to background ratios.}\label{fig:SynEx3}
\end{figure}

\subsection{Synthetic experiment corresponding to Fig. 1(e) of the main paper}\label{sec:resolution}
For quantifying the resolution of MUSICAL and obtaining
the plot shown in Fig. 1(e) of the main paper, we considered
a sample containing two lines of 500 nm each. Each line
contains 51 emitters placed uniformly at a separation of 10
nm. The distance between the two lines is characterized by
$\Delta x$, as shown in Fig. 1(e) of the main paper. The simulation
of blinking is discussed in section III-E. One image stack
of 1000 frames is generated for each value of $\Delta x$, which
varies between 20 nm and 60 nm. The image stacks are
provided as tiff files with the name SynPairDelX. The file with name SynPairDelX500Angstrom represents the image stack with $\Delta x = 500$ \AA $= 50$ nm. The image stacks are noise free.

\subsection{Simulation of blinking in synthetic examples}\label{sec:blinking}

SynPairDelX image stacks, SynEx1, SynEx2, SynEx3, SynEx4,
SynPeriod, SynFork1 and SynFork2 are generated assuming
Poisson blinking model, where the length of switching times
$t_{{\rm on}}$ and $t_{{\rm off}}$ are computed using a Poisson distribution with the
average on time $\tau_{{\rm on}}$ and the average off time $\tau_{{\rm off}}$, respectively. We have used $\tau_{{\rm on}}=0.3\;{\rm ms},\tau_{{\rm off}}=0.7\;{\rm ms}$ for SynEx1
and SynPeriod; $\tau_{{\rm on}}=3\;{\rm ms},\tau_{{\rm off}}=7\;{\rm ms}$ for SynEx2,
SynEx3, and SynEx4; $\tau_{{\rm on}}=2\;{\rm ms},\tau_{{\rm off}}=998\;{\rm ms}$ for
SynFork1; $\tau_{{\rm on}}=0.2\;{\rm ms},\tau_{{\rm off}}=999.8\;{\rm ms}$ for SynFork2; and $\tau_{{\rm on}}=0.3\;{\rm ms},\tau_{{\rm off}}=999.7\;{\rm ms}$ for SynPairDelX stacks. The photon emission rate is assumed to be 100,000 photons/sec.

\subsection{Simulation of noise in SynEx2, SynEx3, SynEx4, {SynPeriod, }SynFork1, and SynFork2}\label{sec:noise}

For the example SynEx2 used in section \ref{sec:comparison}, we add camera's offset value of 100 and add Poisson noise. In order to achieve a desired SBR, we scale the intensities in SynEx2 between 0 and 100(SBR-1) and add the imaging system's offset value of 100 to the scaled intensities. Then Poisson noise is added to each intensity value. Similarly, the offset and Poisson noise is added to SynEx3, which is used in sections \ref{sec:FWHM} and \ref{sec:highpass}. Offset value of 100, SBR value of 10, and Poisson noise is used for SynFork1 and SynFork2 with.

A fourth synthetic example, SynEx4, has exactly the same details as SynEx2, except that the image stack of SynEx4 has 1000 frames. Then, an image stack, BckEx4, captured for a random region in the actual sCMOS sensor without any laser excitation is fused with the SynEx4 to form a new image stack SynEx4Bck as below
\begin{equation}\label{eq:SynEx4Bck}
{\rm SynEx4Bck}(p,k) = 100 \frac{{\rm SynEx4}(p,k)}{\max\limits_{p',k'} {\rm SynEx4}(p',k')}+{\rm BckEx4}(p,k)
\end{equation}
where $p,k$ denote a pixel and a frame, respectively. Here, 100 is used since the average intensity value of the image stack BckEx4 is 100. Thus, SynEx4Bck has signal to background ratio (SBR) is 2. Only SynEx4Bck is used. Results for this example appear in section \ref{sec:artifacts}.

\begin{table*}
\centering %
\caption{Details of samples are provided here. $^{*}$ Number of washes indicate
the number of times the sample is washed before introducing the imaging
buffer.}
\label{tab:samples}
\begin{tabular}{|p{3.8cm}||p{3.9cm}|p{2cm}|p{1.4cm}|p{1.4cm}|p{2cm}|}
\hline
{Sample}  & Dye  & No. of washes$^{*}$ & No. of frames & Frame rate (fps) & Laser Intensity (W/cm$^2$)\tabularnewline
\hline
\hline
In-vitro Sample 1  & Phalloidin Atto 565  & 3  & 10,000  & 200  & 205.6\tabularnewline
\hline
In-vitro Sample 2  & Phalloidin Atto 565, tetraspeck beads (40 nm diameter) & 5  & 10,000  & 200  & 10.3, 40.2, 205.6\tabularnewline
\hline
In-vitro Sample 3  & Phalloidin Atto 565  & 0  & 20,000  & 100  & 0.93, 10.3, 205.6 \tabularnewline
\hline
\hline
In-vivo microtubules Sample 1 & Lifeact GFP & not applicable & 1,000 & 200 & 25 \tabularnewline
 \hline
In-vivo microtubules Sample 2 & Lifeact GFP & not applicable & 49 & 1,000 & 25 \tabularnewline
\hline
In-vivo F-actin & Lifeact GFP & not applicable & 100 & 1,000 & 25 \tabularnewline
\hline
\end{tabular}
\end{table*}

\subsection{Simulation of synthetic example SynSTORM}\label{sec:SynSTORM}
Image stacks for the SynSTORM are simulated using Matlab
code testSTORM, provided by the authors of \cite{S-rees2012blind}. Photon
emission rate of 100,000 photons/sec, average time of 50,000
seconds before bleaching occurs, and emission wavelength
of 510 nm are used. The parameters of the imaging system
match the experimental system. No background is assumed
(background set to 0). The image stacks contain 3000 frames
each, taken at 20 frames per second.

\section{Additional in-vivo results}\label{sec:invivo}

In this section, we provide additional experiments and results of MUSICAL for in-vivo samples. The first experiment is a time lapse study on the microtubule sample 1 shown in Fig. 4 of the main paper. The second experiment shows result on in-vivo microtubules (sample 2) with small acquisition time of 1 ms. The third experiment is an example of MUSICAL imaging of in-vivo cortical actin cytoskeleton.

\subsection{In-vivo microtubules sample 1, number of frames and time lapse study}
We use region B of in-vivo microtubules sample 1 shown in Fig. 4 of the main paper to discuss the benefit of reducing the number of frames for in-vivo studies involving dynamic molecules. Fig. \ref{fig:LiveCellsSample1TimeLapse} shows MUSICAL images obtained with different numbers of frames and in different time slots of the imaging time-line. The same lower bound criterion (0.4 times of the 99.9\% of the histogram of MUSICAL intensities) is used for all the images. MUSICAL results in the first two columns of the top panel are significantly sharper than the remaining MUSICAL results. This is because the blurring effect due to the local dynamics in the vicinity of the microtubule can be alleviated by using less frames. Further, the background is relatively less cluttered in MUSICAL images with 49 frames. Higher numbers of frames imply collection of more signal from spatially more spread traces of freely diffusing labeled molecules, and thus a more pronounced background. The bottom pane showing MUSICAL results for increasing number of frames indicates that the structural details reconstructed by MUSICAL start saturating at sufficiently large number of frames. Lastly, we note that the images in the first two columns of the top panel, each using 49 frames, do not have overlapping time windows. Unsurprisingly, they differ from each other in the structural details. This indicates the utility of MUSICAL in time lapse studies, allowing direct visualization of dynamics on small time scales. In summary, using the smallest possible number of time frames allows for sharp, less cluttered, super-resolved images with the potential of time lapse studies.

\begin{figure}
  \centering
  \includegraphics[width=\linewidth]{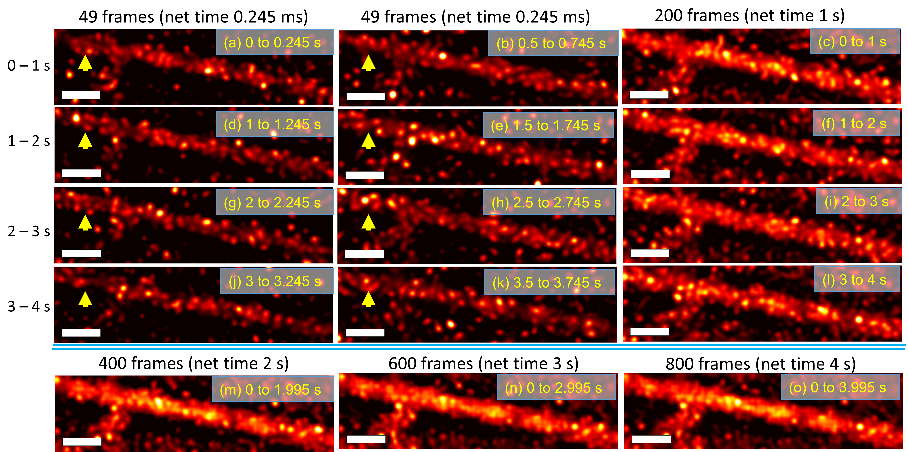}
  \vspace{-7mm}
  \caption{Influence of the number of frames used in MUSICAL for in-vivo microtubules experiment is illustrated here. Scale bars: 500 nm.}\label{fig:LiveCellsSample1TimeLapse}
\end{figure}

\subsection{In-vivo microtubules sample 2, MUSICAL image with 49 ms of temporal data}
Here, we show MUSICAL result for another in-vivo microtubule sample, that is imaged at a frame rate of 1,000 frames per second, i.e. 1 ms per frame. The MUSICAL result for 49 frames is shown in Fig. \ref{fig:LiveCellsSample2}. It is seen that even in this sample, MUSICAL reconstructs sub-100 nm details. The lower bound for this example is chosen heuristically to be 0.30 times the 99.9\% of the histogram of MUSICAL intensities. The lower bound was decreased in the light of the smaller spatial spread of diffusing molecules captured in 49 ms, as opposed to sample 1, where 49 frames correspond to 245 ms.

\begin{figure}
  \centering
  \includegraphics[width=\linewidth]{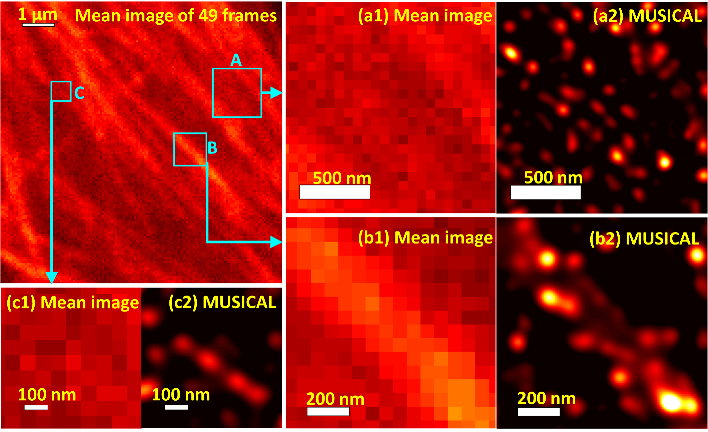}
  \vspace{-7mm}
  \caption{MUSICAL results for in-vivo microtubules sample 2, generated using 49 frames spanning 49 ms of time. Top-left image shows the mean image of 49 frames of the complete sample. (a1-a2) show mean and MUSICAL images of the region A. Similarly for (b1-b2) and (c1-c2).}\label{fig:LiveCellsSample2}
\end{figure}

\begin{figure}
  \centering
  \includegraphics[width=\linewidth]{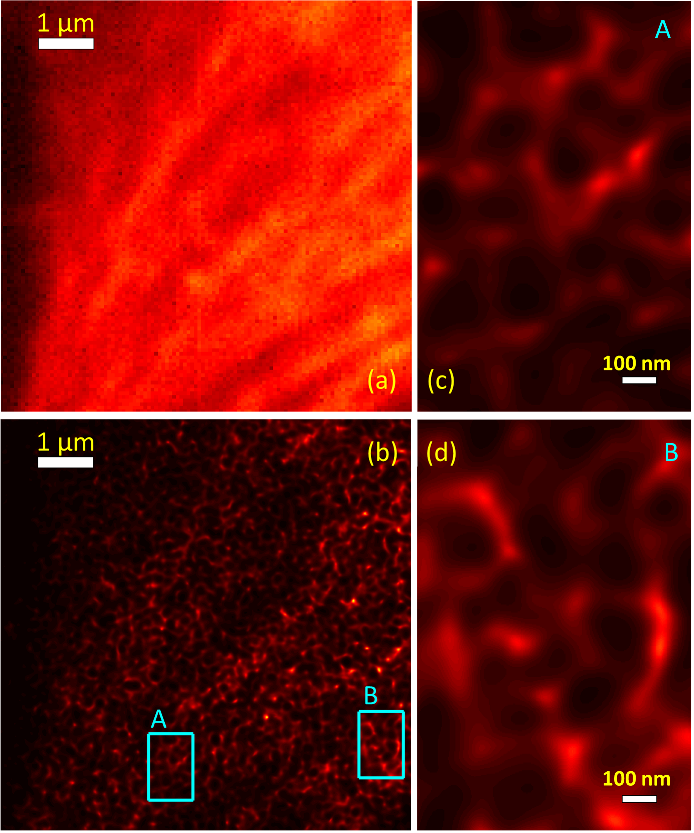}
  \vspace{-7mm}
  \caption{Result of MUSICAL for the in-vivo F-actin experiment is shown here. Mean image of 100 frames (a) has significantly less details than the corresponding MUSICAL image (b). Zoom-ins of regions A and B are shown in (c,d), respectively. Scale bars: (a,b) 1 $\rm \mu$m and (c,d) 100 nm.}\label{fig:LiveCellActin}
\end{figure}

\subsection{In-vivo F-actin}
The in-vivo imaging of actin cytoskeleton was performed in a cell line stably expressing Lifeact-GFP. The whole cytoskeleton is subject to fluorescence labeling. However, due to the TIRF illumination geometry, mainly the cortical actin cytoskeleton was imaged. The cells were illuminated by a 488 nm laser, excitation power of 25 W/cm$^2$, and imaged with 1 ms exposure. The MUSICAL result shown in Fig. \ref{fig:LiveCellActin} was generated using 100 frames only, which corresponds to 100 ms. The results are in good agreement with the known structures of the cortical F-actin mesh \cite{S-bovellan2014cellular,S-charras2006reassembly}. Notably the geometry as well as the mesh size agrees well with Fig. 3(g,h) of \cite{S-bovellan2014cellular} and Fig. 6(d) of \cite{S-charras2006reassembly}.

\section{Additional results and details concerning characterization of MUSICAL}\label{sec:Results}

In this section, we provide additional results on in-vitro actin filaments which serve to supplement the characterization of MUSICAL given in the main paper.

\subsection{Details of in-vitro actin filaments samples}

Three samples of actin filaments were used, two in the main paper
and one in this section. See Table \ref{tab:samples} for a quick
overview of the samples. Main paper showed zoom-in regions A-C of
sample 1 and results of MUSICAL and STORM. In samples 2 and 3, where the same samples are imaged with different
excitation powers, the image stacks have been appropriately shifted
to compensate for the drift of sample between different acquisitions.
This shift enabled better comparison of intensities at sections across
the same features.

\subsection{Periodicity in in-vitro actin filaments}

In order to study the periodicity observed in in-vitro actin filaments, we
performed Fourier analysis of the MUSICAL intensity profiles of lines
A1 and C1 in Fig. 1 of the main paper. The intensity profiles and
their Fourier spectra are shown in Fig. \ref{fig:actin_period_FFT}.
For computing the Fourier amplitudes, the intensity profiles were
mean compensated and fast Fourier transform was computed. As noted
in Fig. \ref{fig:actin_period_FFT}(b), there are two peaks in the
Fourier spectrum for the profile A1 at sampling frequencies corresponding
to 62.5 nm and 75 nm. The other peaks in the Fourier spectrum of A1
are harmonics of these two sampling frequencies. Similarly, there
is one clear peak in the Fourier spectrum for the profile C1 at sampling
frequency corresponding to 65.5 nm and the other peak is the 4th order
harmonic of this frequency.
\begin{figure}
\centering \includegraphics[width=1\linewidth]{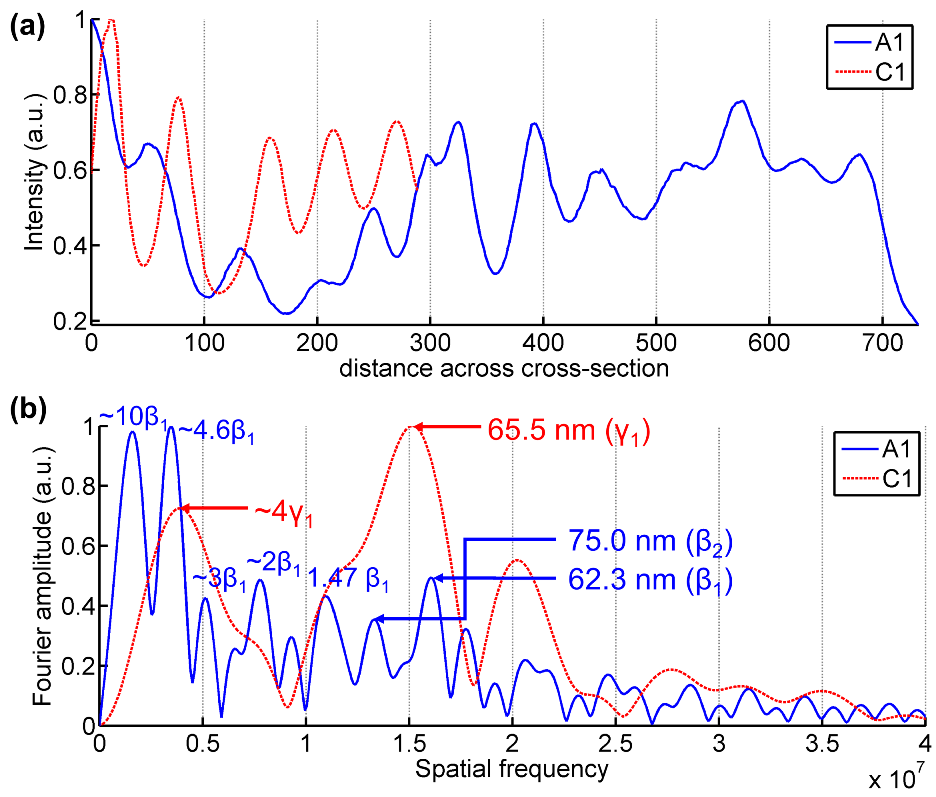}
\vspace{-7mm}
\caption{MUSICAL intensity profiles of lines A1 and C1 in Fig. 1 of the main
paper.}
\label{fig:actin_period_FFT}
\end{figure}
\begin{figure}
  \centering
  \includegraphics[width=\linewidth]{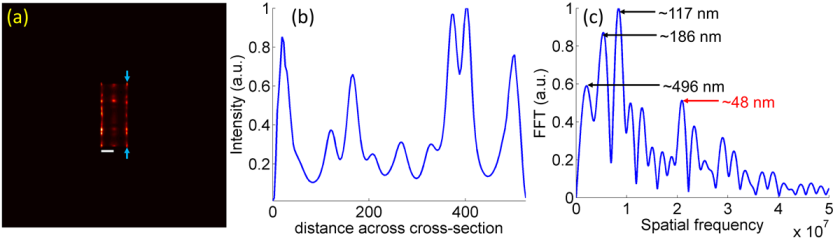}
  \vspace{-7mm}
  \caption{Synthetic example SynPeriod with 33 emitters distributed along 3 lines, which are 100 nm apart. Consecutive emitters along a line are 50 nm apart. MUSICAL result is shown in (a). MUSICAL intensity along the section indicated by the blue arrows in (a) is shown in (b) and its Fourier spectrum is shown in (c). Scale bar: 100 nm.}\label{fig:SynPeriod}
\end{figure}

To further validate the periodicity observed, we simulated a synthetic example SynPeriod which has 33 emitters placed along 3 lines which are 100 nm apart. Each line has 11 emitters uniformly placed along a length of 500 nm, consecutive emitters being 50 nm apart. Example contains 1,000 frames captured at a frame rate of 200 frames per second and the SBR is 100. The MUSICAL result for this example is shown in Fig. \ref{fig:SynPeriod}(a) and the MUSICAL intensity along the third line (indicated by blue arrows) is plotted in (b). The Fourier spectrum is plotted in (c) and shows dominant peaks at about 50 nm or its multiples. This indicates that the peaks observed in Fourier spectra of the in-vitro actin filaments are related to the periodicity of actin filaments.

\subsection{Simulation of forks similar to Fig. 1(b3)}
We use two synthetic examples of fork, SynFork1 and SynFork2, similar to the geometry of fork seen in Fig. 1(b3) of the main paper. Both examples have the same geometry and placement of the emitters. They differ only in their blinking rates such that SynFork2 has much sparser blinking than SynFork1, as discussed in section \ref{sec:Synthetic} before. The fork has two prongs with an angle of $30^\circ$ between them attached to a stem; each prong and the stem are 500 nm long and have uniformly distributed 200 emitters. Image stack contains $K=1000$ frames and has SBR 10. The mean images of the image stack, STORM result, and MUSICAL result are respectively shown in Fig. \ref{fig:SynFork}(a-c) for SynFork1 and Fig. \ref{fig:SynFork}(d-f) for SynFork2. It is seen that STORM can reconstruct the fork with good fidelity only when the blinking is extremely sparse. The nature of STORM result for SynFork1 with speckles between the prongs matches with the STORM result for the fork in Fig. 1(b3).

\begin{figure}
  \centering
  \includegraphics[width=\linewidth]{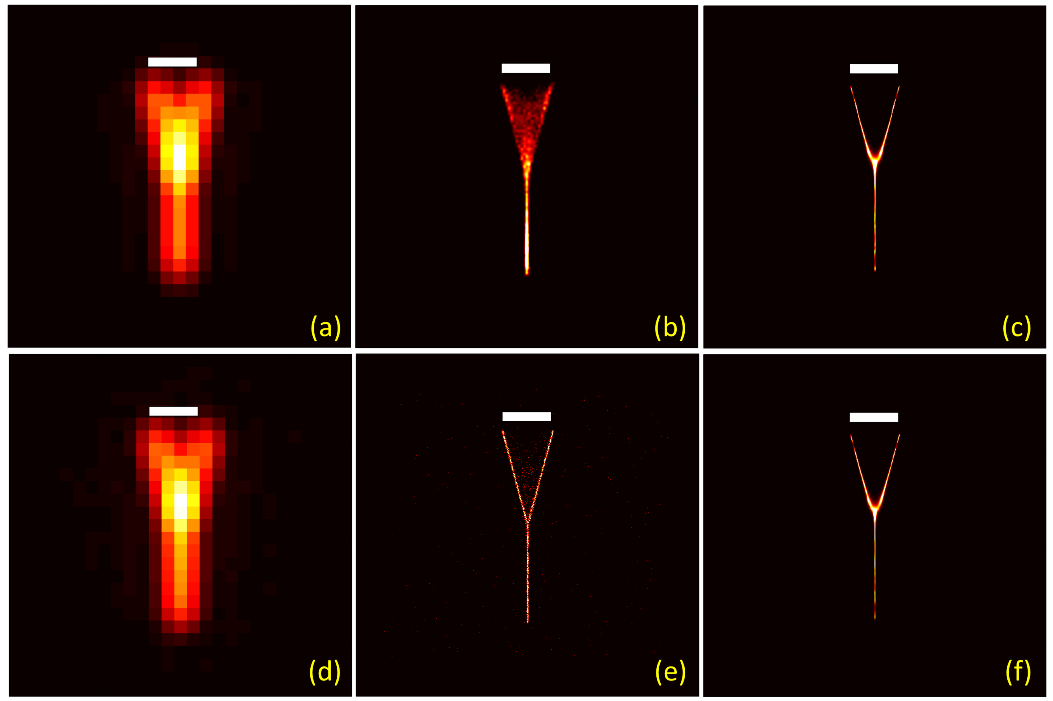}
  \vspace{-7mm}
  \caption{Results for synthetical experiments SynFork1 and SynFork2 which simulate a fork similar to Fig. 1(b3) of the main paper. Top row: SynFork1;Bottom row:SynFork2. (a,d) mean image of 1,000 frames. (b,e) STORM result (c,f) MUSICAL result. Scale bar: 250 nm.}\label{fig:SynFork}
\end{figure}

\subsection{MUSICAL and side lobes}\label{sec:highpass}

To illustrate non-zero side lobes in MUSICAL images, we consider
the region D shown in Fig. \ref{fig:highpass}(a) and present MUSICAL
intensities at 3 sections D1-D3, shown in Fig. \ref{fig:highpass}(b).
In the plots of normalized intensities, shown in Fig. \ref{fig:highpass}(c),
the presence of slide lobes is clearly evident. We note that this
effect is similar to the well-known effect of increased side lobes
when pupil filters such as annular filters are introduced to reduce
the size of the focal spot of a focusing system \cite{S-born1999principles}.
Analogously, the improvement of resolution in MUSICAL as compared
to the original image stack is accompanied with side lobes. However,
as noted in Fig. \ref{fig:highpass}(d), the level of side lobes remains
almost the same for all the three curves, irrespective of their intensities.

Further, we plot the side lobe levels relative to the peak intensity of MUSICAL result ($\alpha=4$) as a function of the SBR for SynEx3 in Fig. \ref{fig:SynEx3}(b). For low values of SBR, the side lobe levels may be quite prominent. We also note that such side lobes appear as well in other results in electromagnetic imaging problem \cite{S-hou2006direct}, although they were not discussed there. In practice, this effect of side lobes can be removed by adjusting
the minimum intensity of the color bar to a small non-zero value commensurate with the side lobe levels.

\begin{figure}
\centering \includegraphics[width=\linewidth]{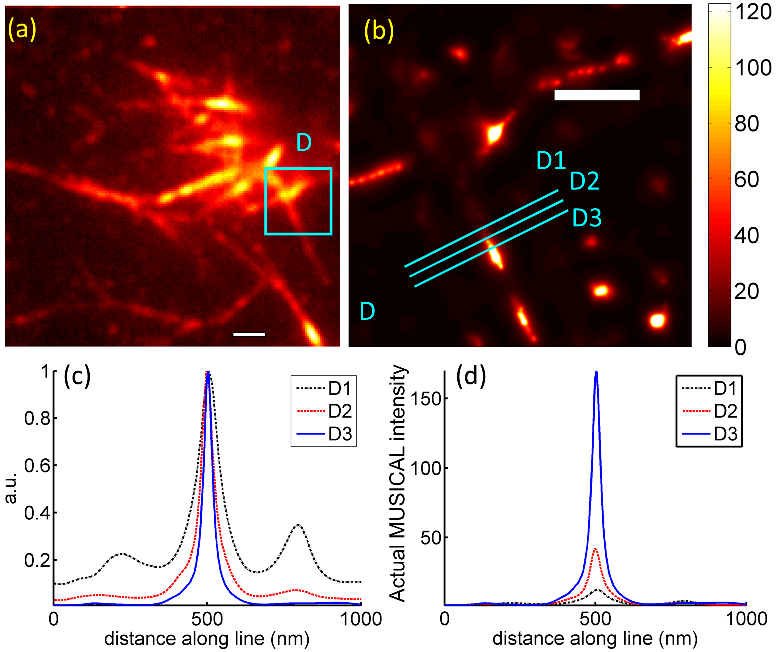}
\vspace{-7mm}
\caption{Mean image of sample 1 and the region D used here are shown in (a). MUSICAL result for region D is shown in
(b). Normalized MUSICAL intensity profiles from sections D1-D3 are
shown in (c) and the corresponding actual intensities are shown in
(d). Scale bars: (a) 1 $\rm \mu$m, (b) 500 nm.}
\label{fig:highpass}
\end{figure}

\subsection{High density of fluorophores in in-vitro sample 2 and at the junction in in-vitro sample 1, zoom-in region A}\label{sec:density}

We identify four regions of $7 \times 7$ pixels, 2 each in in-vitro samples 1 and 2, as shown in Fig. \ref{fig:density}(a,b). The sum of intensities in the measured image stack in each of these regions is plotted as a function of the frame number in Fig. \ref{fig:density}(c). The plotted intensities are an indicator of density of fluorophores in the region since all the remaining experimental factors are the same. The regions R3 and R4 correspond to the branch of actin filaments in zoom-in A of sample 1 and a single actin filament, respectively. It is seen in the inset of Fig. \ref{fig:density}(c) that the intensity in R3 is approximately 1.8 times the intensity in R4, indicating more densely packed fluorophores in R3. Further, as seen in Fig. \ref{fig:density}(c), the intensities in sample 2 (R1 and R2) are significantly larger than in sample 1 (R3 and R4), indicating much higher density of fluorophores in sample 2 than in sample 1. For the same distribution of on and off times, higher density of fluorophores translates to less sparse blinking.

\begin{figure}
  \centering
  \includegraphics[width=0.9\linewidth]{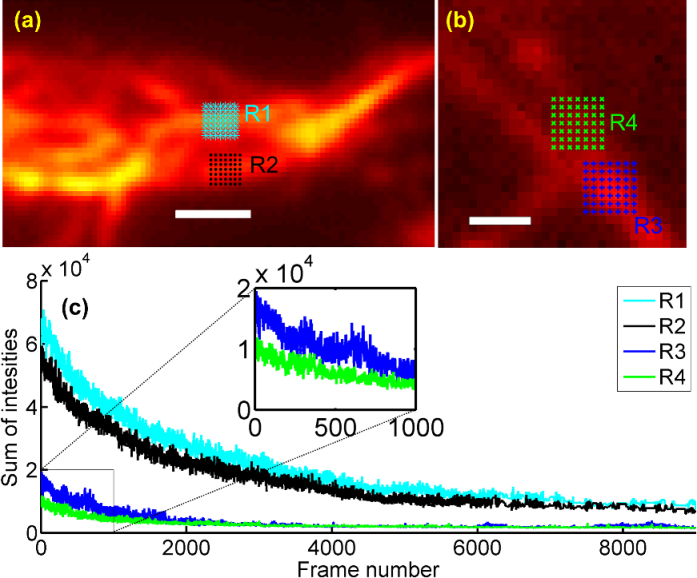}
  \vspace{-3mm}
  \caption{Plots of sum of the measured intensities in different $7\times7$ regions R1 and R2 of sample 2 (a) and R3 and R4 of sample 1 (b) are shown in (c). All the plots correspond to image stacks captured with excitation laser's intensity of 205.6 W/cm$^2$. Scale bars: (a) 1 $\rm \mu$m; (b) 500 nm.}\label{fig:density}
\end{figure}

\subsection{STORM results for in-vitro sample 2}\label{sec:STORM_sample2}

The STORM results for in-vitro sample 2 for the excitation laser's intensities 10.3 W/cm$^2$, 40.2 W/cm$^2$, and 205.6 W/cm$^2$ are given in Fig. \ref{fig:STORM_sample2}. It is seen that there are much less details in STORM results compared to MUSICAL results in Fig. 2 of the main paper. As already inferred in section \ref{sec:density}, the blinking in sample 2 is not too sparse even for 205.6 W/cm$^2$ and thus is not well-suited for STORM. Unsurprisingly, MUSICAL outperforms STORM for this example.

\begin{figure}
  \centering
  \includegraphics[width=\linewidth]{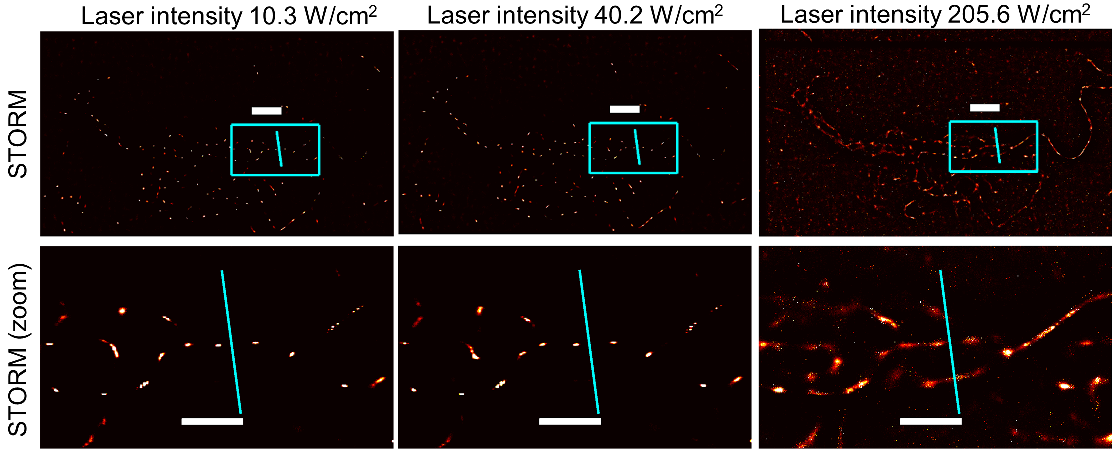}
  \vspace{-7mm}
  \caption{STORM results for sample 2 are shown here. The outline of the zoom-in regions and the section line have the same coordinates as Fig. 2 of the main paper.}\label{fig:STORM_sample2}
\end{figure}

\subsection{Results of in-vitro sample 3 for different excitation powers}

We considered another in-vitro sample, Sample 3, for which image stacks were
acquired at excitation laser powers 0.93 W/cm$^2$, 10.3 W/cm$^2$, and 205.6 W/cm$^2$. The
results are shown in Fig. \ref{fig:power}. The MUSICAL results show
sharper images than the mean images for all the three powers, see
Fig. \ref{fig:power}(a3,b3,c3). Comparison of MUSICAL results at
the cross-sections D1-D3 are shown in Fig. \ref{fig:power}(d). It
is notable that as the power decreases from 205.6 W/cm$^2$ (D3) to 10.3 W/cm$^2$ (D2),
FWHM deteriorates from 38.1 nm to 248.4 nm. However, as the power
decreases from 10.3 W/cm$^2$ to 0.93 W/cm$^2$, the FWHM deteriorates by only 20
nm. The FWHM of mean intensities at the cross-sections D1 and D2 are 320 nm and 336 nm, respectively. Thus, MUSICAL result is slightly better than the unprocessed image stack even when the excitation power is low.

Thus, collecting the results shown for in-vitro Sample 2 in Fig. 2 of the
main paper and the result for sample 3, we form an empirical plot
of power response of MUSICAL shown in Fig. \ref{fig:power}(e). We
note that there are three factors responsible for the observed steep
curve in Fig. \ref{fig:power}(e). As shown in Dempsey et. al
\cite{S-dempsey2011evaluation}, reduction of excitation laser
intensity has two effects that affect the captured signal directly.
First, reduced intensity reduces the number of photons emitted per
switching cycle, thus deteriorating the SNR as well as the SBR of
the image stack. Notably the SBR does not change significantly in the flat regions of the curve in Fig. \ref{fig:power}(e). Second, reduced excitation intensity reduces the off-switching rates and thus increases the duty cycles, which translates to less dynamic variation in blinking or flatter statistics. Further, a third effect is the survival rate of emitters, which reduces with increasing excitation power. We believe that these three effects compete with each other to
result into the empirical plot of Fig. \ref{fig:power}(e). We note here
that since the total acquisition time duration is significantly shorter than
the time needed to reach the equilibrium on-off duty cycle (400-600 seconds
\cite{S-dempsey2011evaluation}), the duty cycle does not reach the
equilibrium value, which is almost the same for all the power values.

\begin{figure}[t]
\centering \includegraphics[width=1\linewidth]{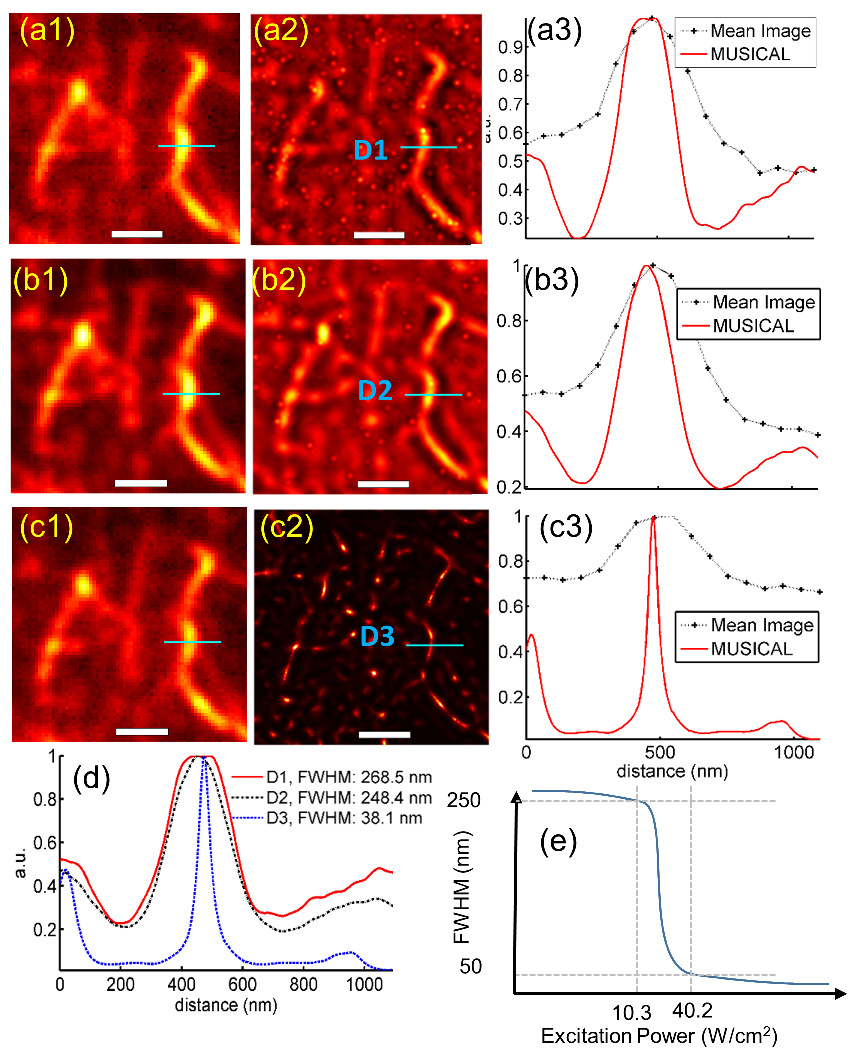}
\vspace{-7mm}
\caption{Actin filaments are imaged using a TIRF microscope and an excitation
laser with power 0.93 W/cm$^2$ (a1-a3, top row), 10.3 W/cm$^2$ (b1-b3, second row),
and 205.6 W/cm$^2$ (c1-c3, third row). The mean images are shown in (a1,b1,c1),
while the MUSICAL results are shown in (a2,b2,c2). Intensities in
the mean image and the MUSICAL image for cross-sections D1-D3 are
compared in (a3,b3,c3), respectively. MUSICAL result of a region
containing Poly-L-lysine only is shown in (d). Comparison of MUSICAL
for cross-sections D1-D3 is shown in (e). Scale bars are 1 micron.
Empirical plot of power versus MUSICAL FWHM is shown in (f).}
\label{fig:power}
\end{figure}

\subsection{MUSICAL and sensitivity to the camera noise}\label{sec:artifacts}

\begin{figure}
\centering \includegraphics[width=1\linewidth]{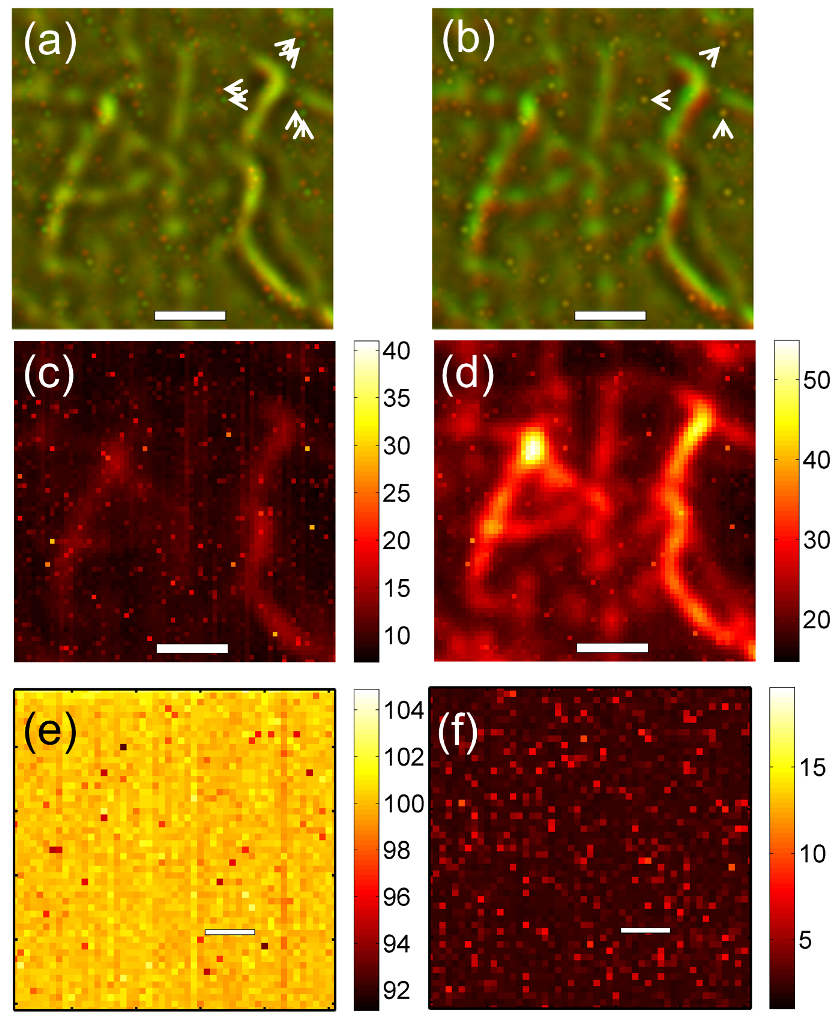}
\vspace{-7mm}
\caption{MUSICAL results for sample 3 acquired using laser intensities 0.93 W/cm$^2$ and 10.3 W/cm$^2$ are fused as red and green channels respectively and shown
here. (a) shows the result with the samples aligned while (b) shows
the fusion of MUSICAL image without sample alignment.(c,d) Standard deviations's map for the measured intensities using laser intensities 0.93 W/cm$^2$ and 10.3 W/cm$^2$, respectively. (e,f) Mean and standard deviations, respectively, of the measured intensities in a random image region of scMOS sensor measured without any laser excitation. Scale bars in (a-d) and 1 micron each and (e,f) are 500 nm each.}
\label{fig:system_sen}
\end{figure}

MUSICAL result for Sample 3 provides another interesting observation
for low powers. Fig. \ref{fig:power}(a2,b2) show some small particle-like artifacts,
with relatively less intensity than the main structures. We discuss
the cause of the appearance of these artifacts. For
this, we fuse the MUSICAL images corresponding to image stacks captured
with 0.93 W/cm$^2$ and 10.3 W/cm$^2$ and show the fused result in Fig. \ref{fig:system_sen}.
Fig. \ref{fig:system_sen}(a) shows the fused images where the MUSICAL
results have been shifted to compensate for the sample drift between
the two acquisitions, whereas Fig. \ref{fig:system_sen}(b) shows
the fusion without compensating for the sample drift. We highlight,
using white arrows, some of the artifact in the fused images.
Notably in Fig. \ref{fig:system_sen}(b), the artifacts are at the
same location but the samples are not aligned, as seen using two distinct
non-overlapping red and green casts in the sample details. On the
other hand, Fig. \ref{fig:system_sen}(a) shows the samples aligned
but two red and green artifacts for each artifact shown in Fig. \ref{fig:system_sen}(b).
In other words, if the samples are aligned the artifacts do not overlap and
if the artifacts are aligned, the samples are shifted.

This indicates that the artifacts are neither sample related nor algorithm
induced, because the particles would have moved with the
sample in either of these cases. Interestingly, the standard deviation maps in Fig. \ref{fig:system_sen}(c,d) of the image stacks show that some background pixels, where the artifacts were observed have larger standard deviations than the other background pixels. Upon further analysis, we found that the dark noise characteristics of the pixels of the sCMOS camera used for measurements are not uniform despite the in-built hot-pixel correction. Particularly, some pixels exhibit significantly larger standard deviations than the rest as seen for a randomly selected region in Fig. \ref{fig:system_sen}(f), although the mean values appear relatively uniform as seen in Fig. \ref{fig:system_sen}(e).

We used the semi-synthetic example SynEx4Bck, which is a fusion of synthetic data and measured dark noise, see eq. (\ref{eq:SynEx4Bck}), to validate the occurrence of such artifacts. The mean and standard deviations of the image stack are shown in Fig. \ref{fig:SynEx4}(a,b) and demonstrate roughly similar behaviour as the mean and standard deviation maps of Sample 3. The SBR is about 1.5, as noted from the mean image in Fig. \ref{fig:SynEx4}(a). The MUSICAL result in Fig. \ref{fig:SynEx4}(c) shows particle-like artifacts in the background in addition to the pronounced effect of side lobes  due to poor SBR. We note that the maximum MUSICAL intensity for SynEx4Bck is about 72. Thus, according to Fig. \ref{fig:SynEx3}(b), side lobes with intensity of about 6.5 are expected, which agrees with the observed values in Fig. \ref{fig:SynEx4}(c).

\begin{figure}
  \centering
  \includegraphics[width=\linewidth]{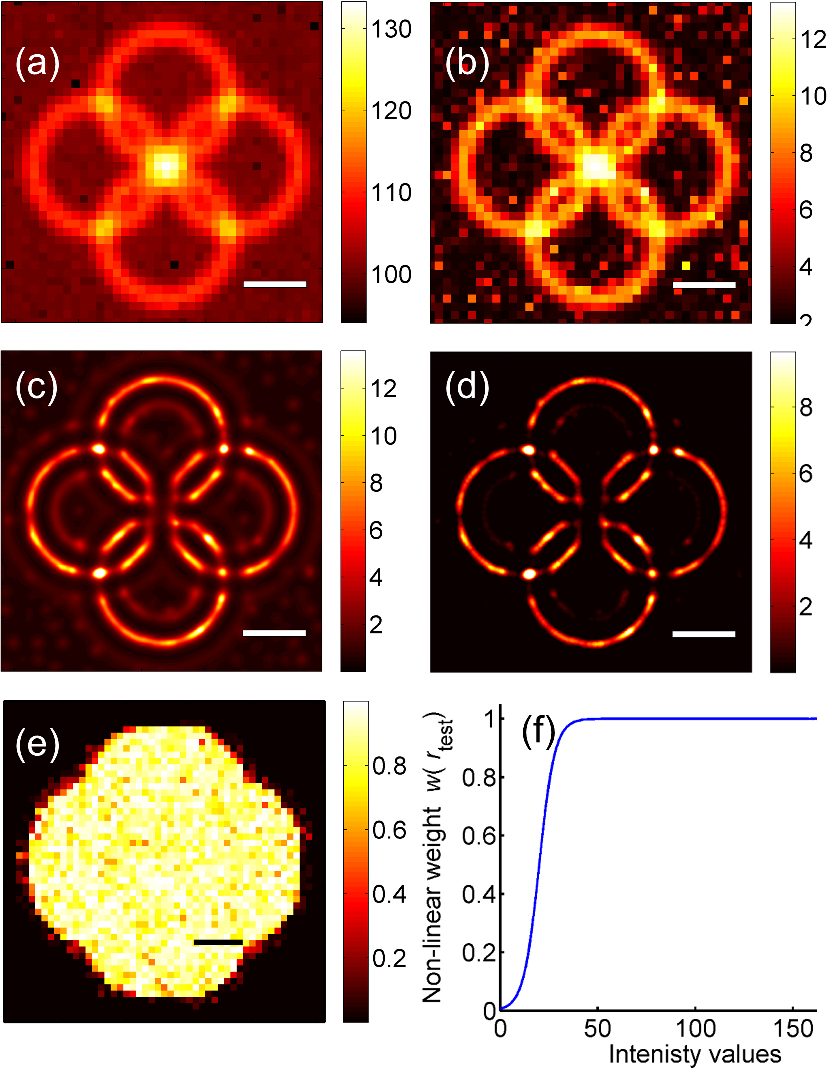}\\
  \caption{Results for synthetic example SynEx3Bck are shown here. (a,b) Mean and standard deviations, respectively, of the image stack. (c) original MUSICAL result (d) MUSICAL results with compensation of noise and suppression of side lobes. (e) Scale map computed using eq. (\ref{eq:scale}). (f) Non-linear weighing function computed using eq. (\ref{eq:sidelobe}). Scale bars are 500 nm each.}\label{fig:SynEx4}
\end{figure}

\subsection{Compensating for known noise pattern of the camera and suppressing side lobes}
We demonstrate here a method each for compensating for known camera noise pattern and for suppressing side lobes. While these approaches are not needed when the SBR is sufficiently high, they may be useful for low SBR.

If an image stack of the dark noise of the same image region can be captured, its singular value decomposition can be used to suppress these artifacts provided that they are not too close to the the features of interest. Suppose the sliding window at a pixel $p$ gives the small image stack ${\bf I}_p$  and noise stack ${\bf N}_p$ and their respective eigenimages are denoted as $\bar u^{\rm meas}_i$ and $\bar u^{\rm noise}_i$, then the net projection on the dark noise space onto the measurement space is given as

\begin{equation}\label{eq:proj}
  d = \frac{\sum\limits_i \sum\limits_j \bar u^{\rm meas}_i \cdot \bar u^{\rm noise}_j}{N^{\rm meas}N^{\rm noise}}
\end{equation}
where $N^{\rm meas}$ and $N^{\rm noise}$ are the number of eigenimages of ${\bf I}_p$ and ${\bf N}_p$, respectively. Then, a scale factor ${\rm scale}(p)$

\begin{equation}\label{eq:scale}
  {\rm scale}(p) = 1-|d|
\end{equation}
can be used to scale the indicator function of MUSICAL. The scale factor for SynEx3Bck is shown in Fig. \ref{fig:SynEx3}(e).

For compensating for the side lobes, we use non-linear weighing of the MUSICAL indicator function, where the value of the indicator function at each test point $\vec r_{\rm test}$ is weighed by $w(\vec r_{\rm test})$ function defined as follows

\begin{equation}\label{eq:sidelobe}
  w(\vec r_{\rm test}) = \frac{1}{1+\exp\Big({-\beta\big({f(\vec r_{\rm test})-f_0}\big)}\Big)}
\end{equation}
where $\beta$ is a control parameter that determines the rate at which the function decays. Here, we choose $\beta=0.25$. Further, $f_0$ is the soft threshold for the intensity. We recommend $f_0$ to be a fraction (we have used 0.3) of the maximum intensity of MUSICAL result. For example, here we have used $I_0=20$ for SynEx3Bck, which is about 0.3 times the maximum MUSICAL intensity. The weights thus calculated for SynEx3 are plotted in Fig. \ref{fig:SynEx3}(f). As seen in Fig. \ref{fig:SynEx3}(f), the indicator function does not change much if its value is more than 40.

For SynEx3, the MUSICAL result after incorporating both ${\rm scale}(p)$ and $w(\vec r_{\rm test})$ is shown in Fig. \ref{fig:SynEx3}(d). Both the artifacts and the side lobes are significantly suppressed in this result as compared to the original MUSICAL image shown in Fig. \ref{fig:SynEx3}(c).

As a further example, we show the effect of $w(\vec r_{\rm test})$ on the MUSICAL image of sample 2 in Fig. 2(a4) of the main paper. We use $\beta=1$ and $f_0=6.85$ which is approximately 0.3 times the maximum MUSICAL intensity. The results are shown in Fig. \ref{fig:sidelobe_correction}. The MUSICAL image with $w(\vec r_{\rm test})$ has a significantly lower background as evident in both Fig. \ref{fig:sidelobe_correction}(b,c).

\begin{figure}
  \centering
  \includegraphics[width=\linewidth]{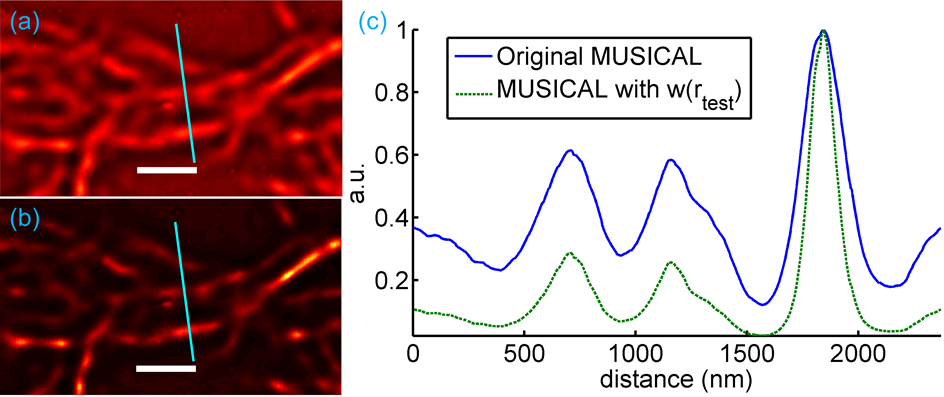}\\
  \caption{Reduction of background using $w(\vec r_{\rm test})$ in the MUSICAL result for sample 2. (a) zoom-in section of MUSICAL result for sample 2, the same as Fig. 2(a4) of the main paper. (b) the same section, but with the incorporation of $w(\vec r_{\rm test})$. (c) comparison of cross-sections.}\label{fig:sidelobe_correction}
\end{figure}

\section{Comparison with other super-resolution techniques}

\label{sec:comparison}

\subsection{Comparison of the computational aspects of super-resolution
techniques}
We present a succinct juxtaposition of the computational
aspects of MUSICAL with the other methods in Table \ref{tab:qualitativeComparison}.
While the number of frames required by MUSICAL is very
small, SOFI and the single iteration ESI have similarly low
requirement; the basic condition for all three is capturing
a statistically significant sample of blinking. On the other
hand, MUSICAL allows any scale of sub-pixelation without
additional iterations unlike those in ESI or the crosscorrelations
in SOFI. MUSICAL scores better than 3B in
computation time but 3B outperforms MUSICAL in image
sharpness as the signal to noise ratio becomes poorer, due
to its iterative rejection of less likely localization candidates.
Further, as compared to SOFI, ESI, as well as 3B, MUSICAL
requires the PSF to be known with sufficient accuracy although
its sensitivity to the error in PSF estimation is reduced by
choosing a sufficiently high value of $\alpha$. The only aspect where
MUSICAL is somewhat heuristic is in the choice of its control
parameter $\sigma_0$. However, the rules of thumb noted in section
II-E are sufficient in most cases. Further, the appearance of
grid artifacts despite using the modified indicator function and
$\alpha \ge 2$ is a good indication of an inappropriate choice of the
value of $\sigma_0$.

\subsection{Details of algorithmic implementation and control parameters
of MUSICAL and other methods}\label{sec:parameters}
Here, we discuss the details of the implementations and
values of control parameters for all the methods used in this
paper for reporting results.

\begin{table*}
\centering %
\caption{Qualitative comparison of super-resolution techniques. $^{\#}$ Limited by blinking statistics. }
\label{tab:qualitativeComparison}
\begin{tabular}{|l||l|l|l|l|l|l|l|}
\hline
Technique  & Instrumentation  & Dye  & Number  & Frame  & Excitation  & Spatial  & Resolution\tabularnewline
{}  & dependence  & dependence  & of frames  & rates (fps)  & Power  & view region & \tabularnewline
\hline
\hline
STED  & Yes  & Yes  & 1  & $\approx$ 10-100  & High  & Unlimited  & $\approx$ 20 nm\tabularnewline
STORM  & No  & Yes  & $\ge$1000  & $\approx$ 10-100  & High  & Unlimited  & 20-30 nm\tabularnewline
3B  & No  & No  & 100-1000  & \#  & Medium  & 100 to 10,000 pixels  & $\approx$ 50 nm\tabularnewline
SOFI  & No  & No  & $\ge$100  & \#  & Medium  & Unlimited  & $\approx$ 100 nm\tabularnewline
ESI  & No  & No  & $\ge$100  & \#  & Medium  & Unlimited  & $\approx$ 80 nm\tabularnewline
MUSICAL  & No  & No  & $\ge$50  & \#  & Medium  & Unlimited  & 30-50 nm\tabularnewline
\hline
\end{tabular}
\end{table*}

\subsubsection{MUSICAL} MUSICAL has been implemented on Matlab
2012b. Parallelization has not been implemented. For
generating the results in the papers, camera's offset value
of 80 is subtracted for experimental images and synthetic
examples for which the signal to background ratio is defined.
Gaussian soft window of size $7\times7$ pixels is used. The standard
deviation of the Gaussian soft window is 3 pixels. $\alpha=4$ is
used. The signal to background ratio of the mean image after
subtraction of the offset was used as the SNR estimate for
selecting $\sigma_0$ as discussed in section \ref{sec:sigma_0}. If the result appeared unsatisfactory, the knee criterion was used for generating the
MUSICAL result. For synthetic examples without noise,
the knee criterion is used. The point spread function used in
MUSICAL is assumed to be the standard Airy disk computed
using the specified emission wavelength, numerical aperture,
magnification, and pixel size.

\subsubsection{STORM} The rainSTORM Matlab code provided by the
authors of \cite{S-rees2012blind} was used for generating the STORM images.
Least squares based fitting of Gaussian function along $x$ and $y$
axes is done independently on the pixels which show non-zero
difference from all immediate neighbors. We chose a region
of interest of 3 pixels, which translates to using 7 pixels along
either direction for fitting Gaussian function. We tried values
4 and 5 for region of interest as well, but found that using 3
as region of interest gave sharper images. Further, we chose
an estimate of the standard deviation of the Gaussian function
as 2.9 pixels, which corresponds to 188.5 nm. However, as
noted in the user guide, the choice of this parameters is
almost inconsequential since the range of standard deviation
of Gaussian function is quite large. The tolerance is set to 0.2,
as recommended in the user guide. The number of iterations
for fitting the Gaussian function is set to 10, which is more
than the recommended value 6.

For Fig. 3(b) of the main paper, the STORM and PALM
results provided with the dataset are used. Suitable colorbar
has been used for better and brighter visualization of these
results.

\subsubsection{deconSTORM} Matlab implementation of deconSTROM
provided by the authors of \cite{S-mukamel2012statistical} on the project webpage \cite{S-deconSTORM} is
used. Default values were used for all the control parameters
except the parameter $\beta$, i.e. the probability of an inactive
emitter will become active in the next frame. Relevant optical
parameters of the experiment were provided. Different values
in the range $\beta \in [6.5 \times 10^{-5}, 0.65]$ have were attempted and
the best result was reported. $\beta \sim 6.5 \times 10^{-5}$ suited the tubulin
long sequence data in Data-SMLM because of very sparse
blinking. $\beta \sim 0.65$ suited the tubulin high density data in Data-
SMLM and SynEx2 due to non-sparse blinking. $\beta \sim 0.0065$
was used for the remaining data. Sub-pixelation by a factor of
8 was used.

\subsubsection{3B} The 32-bit ImageJ plugin for 3B provided by the
authors of \cite{S-cox2012bayesian} was used for 3B computation. The default
hardcoded values of the control parameters were used. Relevant
optical parameters of the experiment were provided. For
generating the 3B image using the localization results of 3B,
reconstructed pixel size of 3.25 nm (sup-pixelation by a factor
of 20) and reconstruction blur PSF of 30 nm was used. Results
in Fig. 4(a-c) of the main paper are obtained after 50 iterations
and results in Fig. 4(d,e) of the main paper are obtained after
65 iterations.

\subsubsection{ESI} Fiji plugin provided by the authors of \cite{S-ESI} was
used for ESI computation. For the results generated in Fig. 4
of the main paper, two iterations of order 2 each were run. The
number of images binned in the first iteration was a factor of
the number of frames $K$ closest to the $\sqrt{K}$. Number of bins
for entropy was set as 100. Only 1 iteration of ESI was used to
generate the result. Logarithmic scale was used for Fig. 4(d,e)
of the main paper.

\subsubsection{SOFI} Matlab code provided by Dr. Joerg Enderlein for
the article \cite{S-dertinger2010achieving} was used for SOFI. Cumulants of order 5 were
used for Fig. 4(a-c) of the main paper and order 6 were used
for Fig. 4(d,e) of the main paper. Sub-pixelation by a factor
of 5 was used.

\subsubsection{Color scale used for all super-resolution methods} The
non-linearity of all the super-resolution methods mandates the
use of a heuristic range of intensities for better visualization
of the results. We have clipped the maximum values of the
intensity in the following manner. For a result, the cumulative
histogram is computed using 10000 uniformly sized bins. The
bin at which the cumulative histogram reaches 99.9\% is identified.
The colorbar of the result is clipped at the center of this bin.
Thus, only 0.1\% pixels in the result have intensity higher than
the maximum intensity of the colorbar used for visualization. For
MUSICAL, typically 99.96\% is used instead of 99.9\%.

\subsubsection{Machine used for computation} The results of all the
methods were generated using a Dell Precision T5610 workstation
with 64-bit Intel Xeon CPU E5-2630 v2 2.60 GHz
and 64 GB RAM. The 32-bit ImageJ plugin of 3B \cite{S-3B} was
used, however on the same system. The multicore ImageJ
implementation of ESI \cite{S-ESI} was used. No parallelization was
incorporated for SOFI, deconSTORM, and MUSICAL.

\section{Concept of range and null space of matrix and relationship to singular value decomposition}\label{sec:linear}

Any matrix $\bf A$ of size $m \times n$ can be interpreted as a linear mapping from an $n-$dimensional input space to an $m-$dimensional output space. Depending on the nature of $\bf A$, although the output space is $m-$ dimensional, the outputs may be restricted to a smaller subspace of this $m-$ dimensional space. Such nature of $\bf A$ is characterized by rank$(\bf A)<m$. The actual output subspace of the matrix $\bf A$ is called the range $\bf R$ of the matrix $\bf A$ and the matrix $\bf A$ is said to span the range $\bf R$. The complementary subspace to the range $\bf R$, which is not spanned by the matrix $\bf A$ is called the null space $\bf N$. Consider the following matrix as an example:
\begin{equation}\label{eq:ex_A1}
  {\bf{A}} = \left[ {\begin{array}{*{20}{r}}
1&0&2&0\\
0&1&0&2\\
1&1&2&2
\end{array}} \right]
\end{equation}
It is evident that although the above matrix $\bf A$ maps inputs to a $3-$dimensional space, the mapped outputs lie on a $2-$dimensional plane only. The rank of this matrix $\bf A$ is 2 and the aforementioned plane is the range of $\bf A$. All the points in the $3-$dimensional space except the this plane belong to the null space. However, for the matrix $\bf A$ below, the null space is empty:
\begin{equation}\label{eq:ex_A2}
  {\bf{A}} = \left[ {\begin{array}{*{20}{r}}
1&0&2&0\\
0&1&0&2\\
1&1&1&1
\end{array}} \right]
\end{equation}
In the situation that the input space is smaller than the output space, i.e. $m>n$, the null space is definitely not empty. For example, if points on a line (a one dimensional input space) are linearly mapped to a two-dimensional or three-dimensional output space, the mapped points would still be confined to a line.

Since the range of a matrix $\bf A$ is spanned by the linear mapping $\bf A$, any point in the range can be expressed as a linear combination of columns in $\bf A$. On the other hand, any point in the null space cannot be expressed as a linear combination of the columns of $\bf A$. This implies that the entire null space is orthogonal to the columns of $\bf A$ as well as to the range of matrix $\bf A$.

Any point in an $n-$ dimensional space can be represented as a linear combination of $n$ independent vectors. Such a set of $n$ independent vectors is called the set of basis vectors or simply the basis vectors. Although there are infinitely many such sets, it is often desirable to choose basis vectors with specific properties, such as the Euclidean norm of each vector being one, or the basis vectors being orthogonal to each other, or one of the basis vectors being along the direction of maximum variance for a cluster of measured points. Since MUSICAL exploits the range and the null space and their mutual orthogonality, it is desirable to choose a set of basis vectors which allows for convenient identification of the range and the null space, as well as inherently uses orthogonality. Singular value decomposition fulfils these requirement easily. This is because the singular vectors are mutually orthogonal, form a set of basis vectors, and satisfy the following: $\bf A \bar v_i=\sigma_i \bar u_i$, where $\bar u_i$, $\bar v_i$, and $\sigma_i$ are the $i$th left singular vector, right singular vector, and singular value of the matrix $\bf A$, respectively. If $\sigma_i \neq 0$, it implies that $\bar u_i$ can be represented as a linear combination of the columns in $\bf A$, and thus $\bar u_i$ lies in the range. Similarly, if $\sigma_i=0$, $\bar u_i$ cannot be represented as a linear combination of the columns in $\bf A$, and thus belongs to the null space.

\end{document}